\newif\ifdraft
\draftfalse
\newif\ifanon
\anonfalse
\newif\ifappendix
\appendixfalse

\ifanon
  \documentclass[acmsmall,screen,review,nonacm,anonymous]{acmart}
\else
  \documentclass[acmsmall,screen]{acmart}
\fi

\setcopyright{rightsretained}
\acmDOI{10.1145/3689735}
\acmYear{2024}
\acmJournal{PACMPL}
\acmVolume{8}
\acmNumber{OOPSLA2}
\acmArticle{295}
\acmMonth{10}
\acmSubmissionID{oopslab24main-p302-p}
\received{2024-04-05}
\received[accepted]{2024-08-18}


\usepackage{pervasives}

\begin{document}

\title{Knowledge Transfer from High-Resource to Low-Resource Programming Languages for Code LLMs}

\author{Federico Cassano}
\orcid{0000-0002-9318-7454}
\affiliation{%
  \institution{Northeastern University}
  \city{Boston}
  \country{USA}
}
\email{cassano.f@northeastern.edu}

\author{John Gouwar}
\orcid{0000-0003-0494-7245}
\affiliation{%
  \institution{Northeastern University}
  \city{Boston}
  \country{USA}
}
\email{gouwar.j@northeastern.edu}

\author{Francesca Lucchetti}
\orcid{0009-0002-5837-6097}
\affiliation{%
  \institution{Northeastern University}
  \city{Boston}
  \country{USA}
}
\email{lucchetti.f@northeastern.edu}

\author{Claire Schlesinger}
\orcid{0009-0000-2533-1242}
\affiliation{%
  \institution{Northeastern University}
  \city{Boston}
  \country{USA}
}
\email{schlesinger.c@northeastern.edu}

\author{Anders Freeman}
\orcid{0009-0005-1904-6193}
\affiliation{%
  \institution{Wellesley College}
  \city{Wellesley}
  \country{USA}
}
\email{af103@wellesley.edu}

\author{Carolyn Jane Anderson}
\orcid{0000-0001-5717-4210}
\affiliation{%
  \institution{Wellesley College}
  \city{Wellesley}
  \country{USA}
}
\email{ca101@wellesley.edu}

\author{Molly Q Feldman}
\orcid{0000-0002-5222-7720}
\affiliation{%
  \institution{Oberlin College}
  \city{Oberlin}
  \country{USA}
}
\email{mfeldman@oberlin.edu}

\author{Michael Greenberg}
\orcid{0000-0003-0014-7670}
\affiliation{%
  \institution{Stevens Institute of Technology}
  \city{Hoboken}
  \country{USA}
}
\email{michael@greenberg.science}

\author{Abhinav Jangda}
\orcid{0000-0002-4849-6776}
\affiliation{%
  \institution{Microsoft Research}
  \city{Redmond}
  \country{USA}
}
\email{ajangda@microsoft.com}

\author{Arjun Guha}
\orcid{0000-0002-7493-3271}
\affiliation{%
  \institution{Northeastern University}
  \city{Northeastern}
  \country{USA}
}
\affiliation{%
  \institution{Roblox}
  \city{San Mateo}
  \country{USA}
}
\email{a.guha@northeastern.edu}

\makeatletter
  \gdef\shortauthors{Cassano, Gouwar, Lucchetti, Schlesinger, Freeman, Anderson, Feldman, Greenberg, Jangda, and Guha}
\makeatother

\begin{abstract}
Over the past few years, Large Language Models of Code (Code LLMs) have started to have a significant impact on programming practice. Code LLMs are also emerging as building blocks for research in programming languages and software engineering.
However, the quality of code produced by a Code LLM varies significantly by programming language.
Code LLMs produce impressive results on \emph{high-resource programming languages} that are well represented in their training data (e.g., Java, Python, or JavaScript), but struggle with \emph{low-resource languages} that have limited training data available (e.g., OCaml, Racket, and several others).

This paper presents an effective approach for boosting the performance of Code LLMs on low-resource languages using semi-synthetic data. 
Our approach, called \system{}, generates high-quality datasets for low-resource languages, which can then be used to fine-tune any pretrained Code LLM.
\system{} translates training data from high-resource languages into training data for low-resource languages in the following way.
1)~We use a Code LLM to synthesize unit tests for commented code from a high-resource source language, filtering out faulty tests and code with low test coverage.
2)~We use a Code LLM to translate the code from the high-resource source language to a target low-resource language. This gives us a corpus of candidate training data in the target language, but many of these translations are wrong.
3)~We use a lightweight compiler to compile the test cases generated in (1) from the source language to the target language, which allows us to filter our obviously wrong translations. The result is a training corpus in the target low-resource language where all items have been validated with test cases. 
We apply this approach to generate tens of thousands of new, validated training items for five low-resource languages: Julia, Lua, OCaml, R, and Racket, using Python as the source high-resource language.
Furthermore,  we use an open Code LLM (StarCoderBase) with open training data (The Stack), which allows us to
decontaminate benchmarks, train models without violating licenses, and run experiments that could not otherwise be done.

Using datasets generated with \system{}, we present fine-tuned versions of StarCoderBase and Code Llama for Julia, Lua, OCaml, R, and Racket that outperform other fine-tunes of these base models on the natural language to code task. We also present Racket fine-tunes for two very recent models, DeepSeek Coder and StarCoder2, to show that \system{} continues to outperform other fine-tuning approaches for low-resource languages. The \system{} approach is easy to apply to new languages, and is significantly more efficient and effective than alternatives such as training longer.
\end{abstract}

\maketitle

\section{Introduction}
Large Language Models of Code (Code LLMs) are starting to have a significant impact on both professional programmers and research in programming languages and software engineering.
GitHub Copilot is just one of several popular tools powered by Code LLMs~\cite{github-copilot,codewhisperer,tabnine}.
Moreover, Code LLMs are also emerging as a building block for research~\cite{schafer:testpilot,bareissCodeGenerationTools2022,murali2023codecompose,nam2023inide,ross23assistant,lemieux:codamosa,xia:universal-fuzzing,joshi:ring,Phung2023GeneratingHF,First23fse,chen:drb-ml}.
However, the quality of code produced by a Code LLM varies significantly by programming language.
Models are most impressive at producing code in high-resource programming languages such as Python, JavaScript, and Java, but struggle in low-resource languages, such as Racket and OCaml~\cite{cassano:multipl-e,athiwaratkun:mbxp,qinkai:codegeex}.
This puts programmers who rely on these languages at a disadvantage, since they do not receive the same benefits that Code LLMs can deliver for high-resource languages~\citep{ziegler_productivity_2022,murali2023codecompose}.

\begin{figure}
    \centering
    \includegraphics[width=0.83\textwidth]{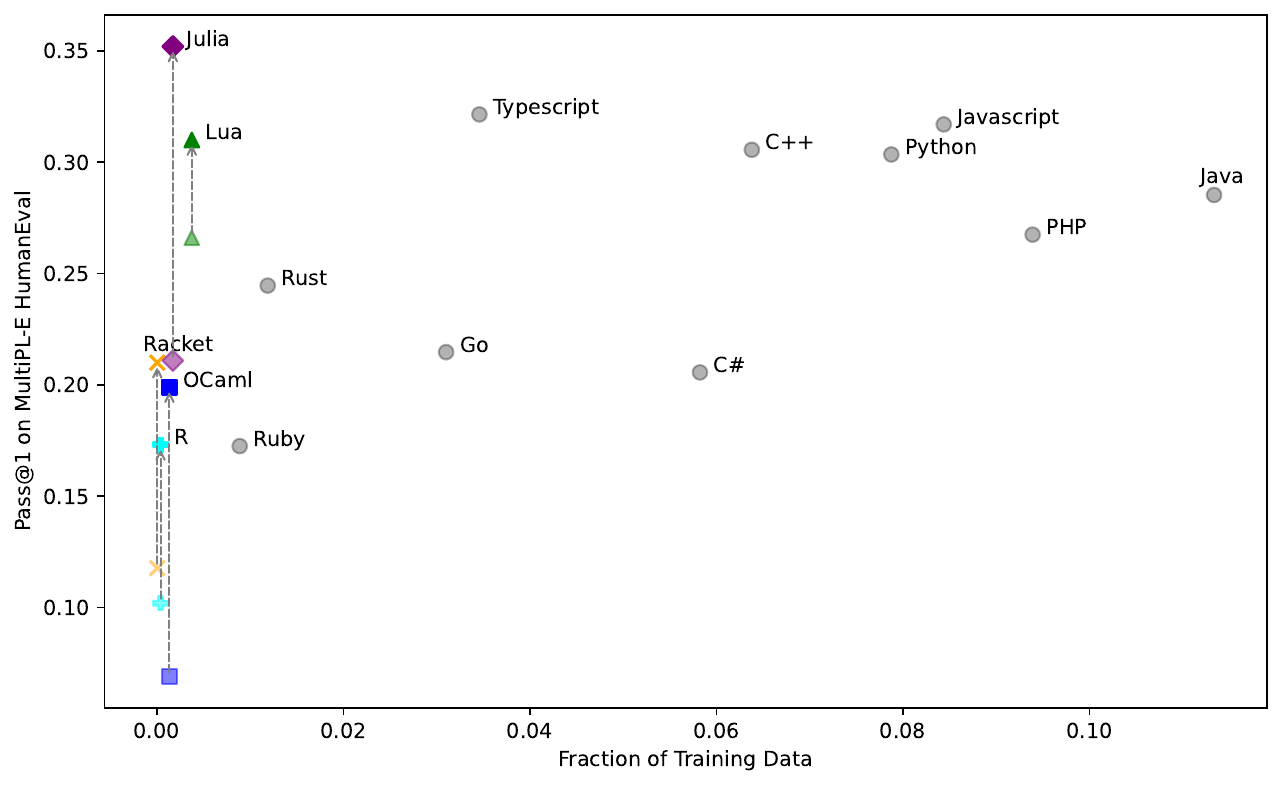}
    \caption{The performance of StarCoderBase-15B on several languages supported by the MultiPL-E benchmark for Code LLMs, plotted against their proportion of the model's training data.
    Using \system{}, this paper significantly improves how StarCoderBase-15B performs on several low-resource languages, as shown by the arrows. The bottom of each arrow indicates how the base model performs, and the arrowheads indicate performance after fine-tuning with \system. We also show significant improvement on other LLMs (\cref{evaluation}).}
    \label{fig:perf_by_freq}
\Description{}
\end{figure}

The key issue is that the performance of Code LLMs depends on the amount of language data available for training. For example, \emph{The Stack}, which is the training set for several contemporary Code LLMs, has 64GB of Python, but only around 1GB of OCaml and 0.5GB of Scheme/Racket~\cite{kocetkov:the-stack}.\footnote{This is the volume of data that remains after files are deduplication for training~\cite{li:starcoder}.} As \cref{fig:perf_by_freq} shows, the performance of \emph{StarCoderBase}, an open Code LLM trained on The Stack, generally increases as the training data for the language increases. 

Our goal in this paper is to investigate methods to further train, or \emph{fine-tune}, pretrained Code LLMs to improve their performance on low-resource languages.
The obvious approach is to try to find more data, but for a low-resource language, more data is hard to find by definition.
For example, The Stack already includes all permissively licensed code for 358 programming languages from GitHub as of 2022, and
GitHub is by far the largest repository of open-source code \cite{github-largest}.\footnote{The Stack deliberately excludes copyleft licenses and unlicensed code.}
An alternative is to train longer (i.e, for more epochs) on existing data.
However, in this paper, we show that \emph{training longer on several low-resource languages is not only inefficient, but can actually hurt performance}~(\cref{just-train-more}).
Another alternative is to train models on synthetic data, which is data generated by an LLM itself. These synthetic fine-tuning datasets are effective for high-resource programming languages~\cite{wang:self-instruct,ziyang:wizard-coder}. However, we show that \emph{synthetic data does not work for low-resource languages} for the intuitive reason that LLMs generate poor quality programs in low-resource languages (\cref{use-self-instruct}).

\paragraph{Our approach}
In this paper, we present a new and effective approach for fine-tuning Code LLMs for low-resource programming languages that is based on generating \emph{semi-synthetic training data}.
Our approach relies on several key ingredients.
1)~The large volume of training data for high-resource programming languages includes a lot of well-documented code; 2)~Code LLMs are effective and efficient unit test generators, and we can check that generated tests are valid~\cite{schafer:testpilot}; 3)~We can compile many unit tests to a low-resource language with a simple compiler~\cite{roziere:transcoder-st,athiwaratkun:mbxp,cassano:multipl-e}; 4)~Code LLMs can translate code from one language to another, and although these translations may be faulty, we can filter them with the aforementioned tests, retry until tests pass, and engineer a prompt to increase the likelihood of a successful translation. Putting these four ideas together, we develop a pipeline for \emph{transferring training data across multiple programming languages} that we call \system. 

\Cref{high-level-figure} gives a high-level overview of how \system{} produces high-quality training data for a low-resource programming language. We use an LLM to translate code from a high-resource language (\circnum{1}) to a target low-resource languages (\circnum{2}). However, this translation is  unreliable by definition: LLMs are bad at producing code in low-resource programming languages. But, we can leverage the stochasticity of LLM generation to produce several candidate translations for each item and filter out the faulty translations with synthesized test cases (\circnum{4}). We cannot generate tests directly from the low-resource code (since it is likely to be faulty). Instead, we generate and validate tests in the high-resource language (\circnum{3}), and then compile these tests to the low-resource language (\circnum{4}). The composition of these steps gives training data in the low-resource that passes compiled tests that also pass in the high-resource language.

The training corpora of high-resource programming languages (e.g., Python) are enormous, so we use aggressive heuristic filters to build a corpus of ``high-quality'' functions (\circnum{1}) before attempting any LLM translation. For these functions, we find that the LLM generates tests reliably (\circnum{3}), but some effort is needed to get high test coverage. Translating functions (\circnum{2}) is simply LLM prompting, but requires some lightweight type inference to be effective when the target language is typed. Finally, to compile tests to the low-resource language (\circnum{4}), we build on an existing toolchain (\cite{cassano:multipl-e}), adding support for OCaml, new prompting formats, and support for an LLM-translation task that is three orders of magnitude larger than what it originally supported.

\begin{figure}
\begin{tikzpicture}[
    font=\footnotesize,
    node distance=2cm and 3cm,
    box/.style={rectangle, draw, rounded corners, minimum width=3cm, minimum height=1.1cm, text centered,text width=1.2in, align=center},
    arrow/.style={-{Latex}}
]

% Nodes
\node[box] (highPL) {\circnum{1}~Function in a high-resource PL};
\node[box, below=of highPL] (testCases) {\circnum{3}~Test cases in the high-resource PL};
\node[box, right=of highPL] (lowPL) {\circnum{2}~Function in a low-resource PL};
\node[box, below=of lowPL] (translatedTestCases) {\circnum{4}~ Test cases in the low-resource PL};

% Arrows
\draw[arrow,orange,thick,Latex-Latex] (highPL) -- node[midway, left, text width=1in, align=center,black] {LLM generated; execution-validated; coverage checked} (testCases);
\draw[arrow,dashed,thick,red,] (highPL) -- node[midway, above,black] {LLM translated} (lowPL);
\draw[arrow] (translatedTestCases)  -- node[midway, right] {Tested} (lowPL);
\draw[arrow] (testCases) -- node[midway, below] {Compiled (no LLM)} (translatedTestCases);
\end{tikzpicture}
\caption{A high-level overview of how \system{} produces high-quality training data for a low-resource programming language. We use a Code LLM to translate a function from a high resource language (\circnum{1}) to a low-resource language (\circnum{2}). The translated code is likely to be wrong, since LLMs perform poorly on low-resource languages. However, we filter out bad translations as follows. First, we generate unit tests the original code (\circnum{3}). We execute these tests to ensure they succeed and also check for test coverage. Second, we compile these tests to the low-resource language (\circnum{4}). Finally, we filter the low-resource code (\circnum{2}) using the translated tests (\circnum{4}), only keeping those that pass tests.}
\Description{See caption.}
\label{high-level-figure}
\end{figure}

Using training data generated by \system, \emph{we present fine-tuned Code LLMs that achieve state-of-the-art performance on five low-resource languages: Racket, OCaml, Lua, R, and Julia.} 
We focus primarily on fine-tuning the StarCoder family of Code LLMs~\cite{li:starcoder}. There are StarCoder models available in a variety of sizes, including a 1B parameter model that is lightweight enough to run on CPUs, and a more capable 15B parameter model, which we use as the test generator and language translator for \system. The StarCoder models also have open training data, which allows us to compare \system{} to a baseline of training longer on existing data for low-resource languages.
We also present fine-tuned versions of the Code Llama 34B and 70B~\cite{codellama} models, and Racket fine-tunes of the recently released DeepSeek Coder~\cite{guo:deepseek} and StarCoder2 models~\cite{starcoder2}.

\paragraph{Contributions}

To summarize, we make the following contributions:
\begin{enumerate}

   \item \system{}, an effective approach for generating semi-synthetic data for low-resource programming languages using test-validated translation of high-quality code in high-resource languages.

   \item Efficient fine-tuning datasets for Julia, Lua, OCaml, R, and Racket, comprising tens of thousands of documented and tested functions generated with StarCoderBase-15B.

   \item A dataset of 133,168 Python functions extracted from the Stack, where every function has natural language documentation and a validated set of generated tests with  high coverage. This dataset could be used to generate fine-tuning sets for other programming languages. 
   
   \item Fine-tuned versions of StarCoderBase 1B and 15B for Julia, Lua, OCaml, R, and Racket. For these languages, these fine-tuned models outperform prior fine-tunes of StarCoderBase on the natural language to code task.

   \item Fine-tuned versions of Code Llama 34B and 70B for Julia, Lua, OCaml, R, and Racket. For these languages, these fine-tuned models outperform prior fine-tunes of Code Llama. More significantly, this is an uncommon result where \emph{data generated from a smaller model (StarCoderBase-15B) improves the performance of larger and better models (Code Llama 34B and 70B)}.

   \item Fine-tuned versions of StarCoder2-15B and DeepSeek Coder 33B for Racket that also outperform other fine-tuned models. These are two very recently released models.

   \item A thorough evaluation that includes a)~a comparison of \system{} to the baseline of training further on existing data, b)~an evaluation of the fine-tuning efficiency with \system{}, c)~results on prior multi-language benchmarks~\cite{cassano:multipl-e}, d)~a new multi-language benchmark designed to exercise in-context learning, e)~an evaluation of how generated code adheres to the common Racket programming style, f)~the impact of data deduplication, and g)~the impact of fine-tuning on the Python source data.
   
\end{enumerate}

\section{Background}
\label{background}

In this section, we give a high-level overview of how Code LLMs are trained and evaluated. We use StarCoder as the example, since it is the model that we use for most of our work.

\subsection{Training and Fine-Tuning Large Language Models of Code}

A \emph{large language model (LLM)} is a neural network trained on hundreds of gigabytes or even terabytes of data. Code LLMs are trained on source code (and often natural language documents too), which allows them to generate code from comments, comments from code, more code from code, and so on. LLM training takes significant resources: StarCoderBase was trained on approximately 800GB of code, which took three weeks on a cluster of 512 NVIDIA A100 GPUs.

The only way to build a training set of this scale is to scrape public repositories of code. There are a handful of public training sets that are based on GitHub~\cite{polycoder,bigquery-code-dataset}, and \emph{The Stack}~\cite{kocetkov:the-stack} is a recent example. The Stack v1.2 has 3TB of permissively licensed source code for 358 programming languages. It was constructed in 2022, and has since been used to train several Code LLMs~\cite{allal:santacoder,replit-code,nijkamp2023codegen2}, including StarCoderBase. Specifically, StarCoderBase was trained on a filtered subset of The Stack consisting of 86 programming languages.

\begin{figure}
% NOTE(arjun): Do *not* delete this figure and replace with a reference to the system diagram. This is a fantastic example of a tricky prompt and a benchmark problem that we want to clearly explain.

\begin{subfigure}{\textwidth}

\begin{lstlisting}[style=codeblock, language=Python]
def vowels_count(s):
    """Write a function vowels_count which takes a string representing a word as
    input and returns the number of vowels in the string.  Vowels in this case are
    'a', 'e', 'i', 'o', 'u'. Here, 'y' is also a vowel, but only when it is at the
    end of the given word. 
    """ 
\end{lstlisting}
\caption{Python prompt.}
\end{subfigure}

\vspace{1em}
\begin{subfigure}{\textwidth}
\begin{lstlisting}[style=codeblock, language=ML]
(* Write a function vowels_count which takes a string representing a word as
   input and returns the number of vowels in the string.  Vowels in this case are
   'a', 'e', 'i', 'o', 'u'. Here, 'y' is also a vowel, but only when it is at the
   end of the given word.  *)
let vowels_count (s : str) : int =
\end{lstlisting}
\caption{OCaml prompt.}
\end{subfigure}
\caption{An example prompt from a HumanEval problem and its translation to OCaml, with our extension to MultiPL-E. Not shown are doctests and hidden test cases, which are also translated to OCaml. This particular problem is hard for many LLMs because it alters the strong prior on what vowels are, by saying that \emph{y is a vowel when it is the last letter in a word}.}
\label{benchmark-ex}
\Description{}
\end{figure}

\paragraph{The StarCoder model family}
StarCoder is a family of models that are available at several sizes~\cite{li:starcoder}. The largest and most capable model in the family is called StarCoderBase, which has 15B parameters.
There are smaller versions of StarCoderBase that were trained on exactly the same data. To make use of limited GPU resources, we use the smallest model, StarCoderBase-1B, for most experiments in this paper. However, we also show that our results generalize to StarCoderBase-15B. 
There is also a model in the StarCoder family that is just named StarCoder: it is StarCoderBase-15B specialized to excel at Python.\footnote{StarCoder fine-tunes StarCoderBase-15B on two more epochs of Python data from The Stack.} This paper uses StarCoderBase-15B for translations to low-resource languages, and StarCoder-15B for Python test generation.

\paragraph{The Code Llama model family}

The Code Llama~\cite{codellama} family of models were recently released and perform better than the StarCoder models on common benchmarks.
While the authors state that the training data comes from publicly accessible datasets, they do not disclose the specific datasets used, preventing us from conducting the exhaustive evaluation on Code Llama that we do with StarCoder and its training data. Moreover, the Llama license forbids using model outputs to train non-Llama models, which is why we use StarCoder for data generation. However, we train and evaluate the larger Code Llama models (34B and 70B).

\paragraph{Fine-tuning} 

After training, a model can be further trained, or \emph{fine-tuned}, with significantly fewer resources. For example, there are several fine-tuned versions of StarCoderBase that were trained with a few days of GPU time on a modest amount of data (e.g., \cite{ziyang:wizard-coder,muennighoff:octopack}). Most fine-tuned versions of StarCoderBase are designed to make the model even better at high-resource languages, such as Python. In contrast, this paper presents fine-tuned versions of StarCoderBase that are significantly better at several low-resource languages.

It is common to \emph{distill} data from a larger model (e.g., GPT-4), to fine-tune a smaller model~\cite{ziyang:wizard-coder,wei:magicoder,gunasekar2023textbooks}. However, we show that \system{} can do the reverse: we use data generated from StarCoderBase-15B to fine-tune larger models, CodeLlama-34B and 70B. 
This is a form of weak-to-strong supervision \citep{burns2023weaktostrong},
where a smaller model is used to generate data for training a larger model,
showing the scalability of \system{}.

\subsection{Code LLM Tasks and Benchmarking Code LLMs}
\label{about-multipl-e}

A Code LLM can be prompted to perform a wide variety of tasks, including code translation (e.g., \cite{pan:lost-in-translation}), test generation (e.g., \cite{schafer:testpilot}), code mutation (e.g., \cite{xia:universal-fuzzing}), code editing (e.g., \cite{cassano:canitedit}), and much more. This article focuses on the \emph{natural language to code task} (e.g., \cite{heidorn:english-vhll}) for low-resource programming languages. Making Code LLMs better on other tasks is beyond the scope of this article.

Most Code LLM benchmarks for the natural language to code task, including those we use in this paper, follow the format introduced by the Codex ``HumanEval'' benchmark~\cite{chen2021evaluating}.
Every benchmark problem has two parts: 1)~a prompt for the LLM that has a function signature and a comment, and 2)~a suite of test cases that are not given to the LLM.
Thus each problem is run in two steps: 1)~the LLM generates a function from the prompt, and 2)~the generated function is then tested with the hidden tests, and all tests must pass for the generated code to be considered correct.

The HumanEval benchmark has 164 problems for Python.
However, it is possible to mechanically translate most of these problems to other programming  languages (\cref{benchmark-ex}).
Translating comments and function signatures is straightforward, but some care is needed to introduce types for typed target languages. Translating test cases turns out to be easy as well, since almost all HumanEval test cases are of the form $f(v_\mathit{in}) = v_\mathit{out}$, where $v_\mathit{in}$ and $v_\mathit{out}$ are first-order values.
This is the approach that is taken by MultiPL-E and similar tools
~\cite{cassano:multipl-e,athiwaratkun:mbxp,babelcode} to build polyglot benchmarks for Code LLMs.
This paper utilizes MultiPL-E, which is the only benchmark to date that supports Racket, and we extend it to support OCaml for this paper.

Code LLMs appear to produce higher-quality code when their output is sampled~\cite{chen2021evaluating}. 
Since sampling introduces non-determinism, we must evaluate their output by generating several samples from the same prompt.  The most widely used metric for Code LLM performance is \emph{pass@k}, which is the likelihood that the LLM produces a program that passes all hidden tests at least once from $k$ attempts. Pass@k must be estimated from $n >> k$ samples. When $k=1$ and there are $c$ successes, pass@1 is the same as the pass rate ($c/n$). We use pass@1 as the metric for all our benchmarking experiments, which is common practice. Intuitively, pass@1 measures the ability of the Code LLM to generate a correct solution in a single attempt.

\subsection{Why StarCoder?}

In the rest of this paper, the majority of the work that we present uses the StarCoder family of Code LLMs for the following reasons.
\begin{enumerate}
    \item At the time that we started this work, StarCoder was the best-performing, permissively licensed, open Code LLM available.
    
    \item The StarCoder family includes a fairly small 1B parameter model, which is amenable to experiments on a budget.
    
    \item Although there were better closed-sourced LLMs when we started, they (a)~did not support fine-tuning, (b)~were significantly more expensive to use at the scale of our work, or (c)~prohibited using their outputs to fine-tune other LLMs~\cite{openai2023terms,google2023terms,anthropic2023terms}. 
    
    \item StarCoder remains the only Code LLM with open training data, which we use to correlate model performance with training set size (\cref{fig:perf_by_freq}), evaluate further training (\cref{just-train-more}), and to check that our benchmarks are not in the training data. It would not have been possible to do this work with any other Code LLM.
    
\end{enumerate}

We do fine-tune several newer Code LLMs with StarCoder-generated data to show that our approach remains useful (\cref{evaluation}).

\section{Alternatives to \system}
\label{alternatives}

Before we present the \system{} approach, we consider two simpler alternatives.

\begin{figure}

\begin{subfigure}{0.41\textwidth}
\includegraphics[width=\textwidth]{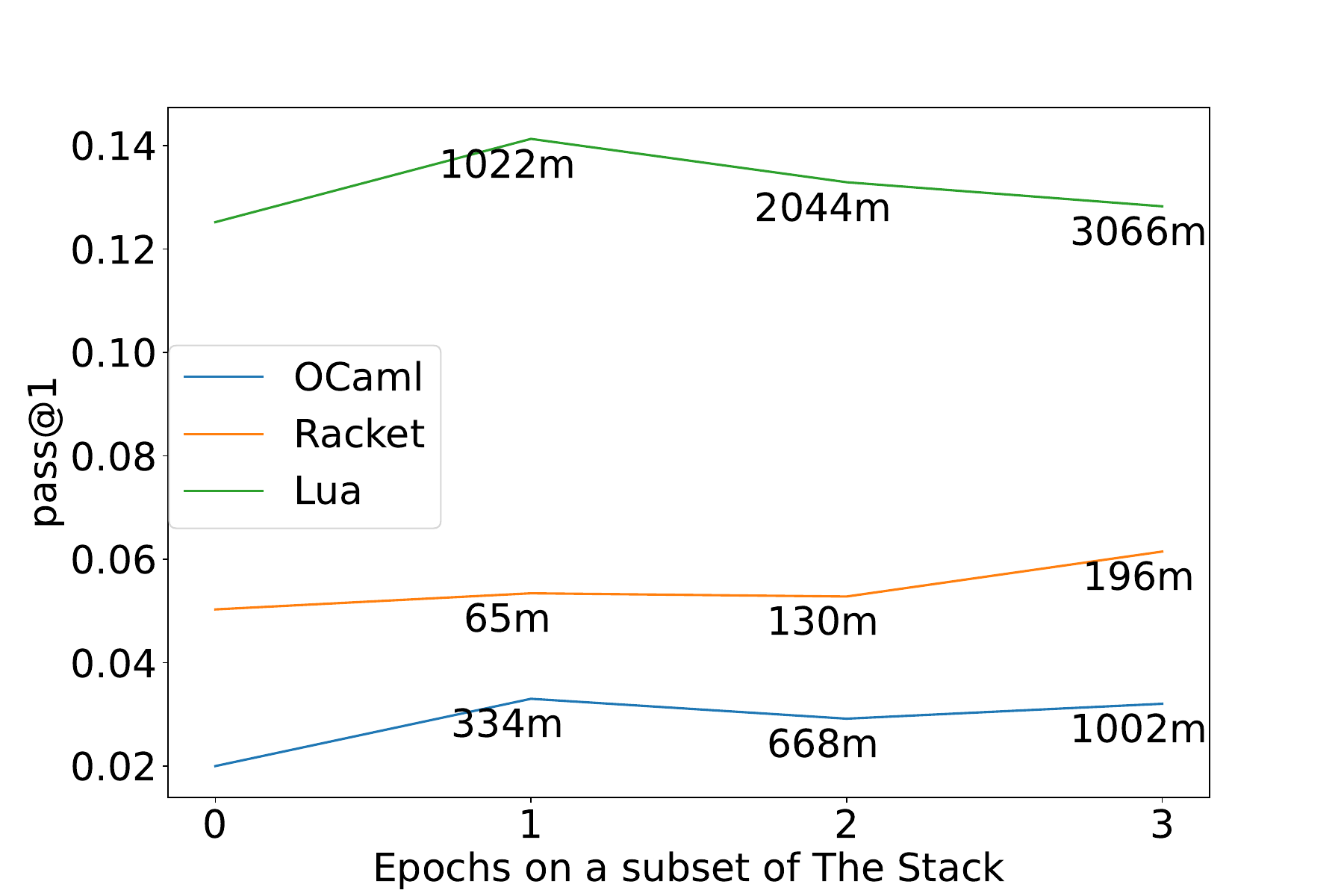}
\caption{Fine-tuning on the complete language \\specific subsets of The Stack.}
\label{strawman-just-train-more-full}
\end{subfigure}
\begin{subfigure}{0.50\textwidth}
\includegraphics[width=\textwidth]{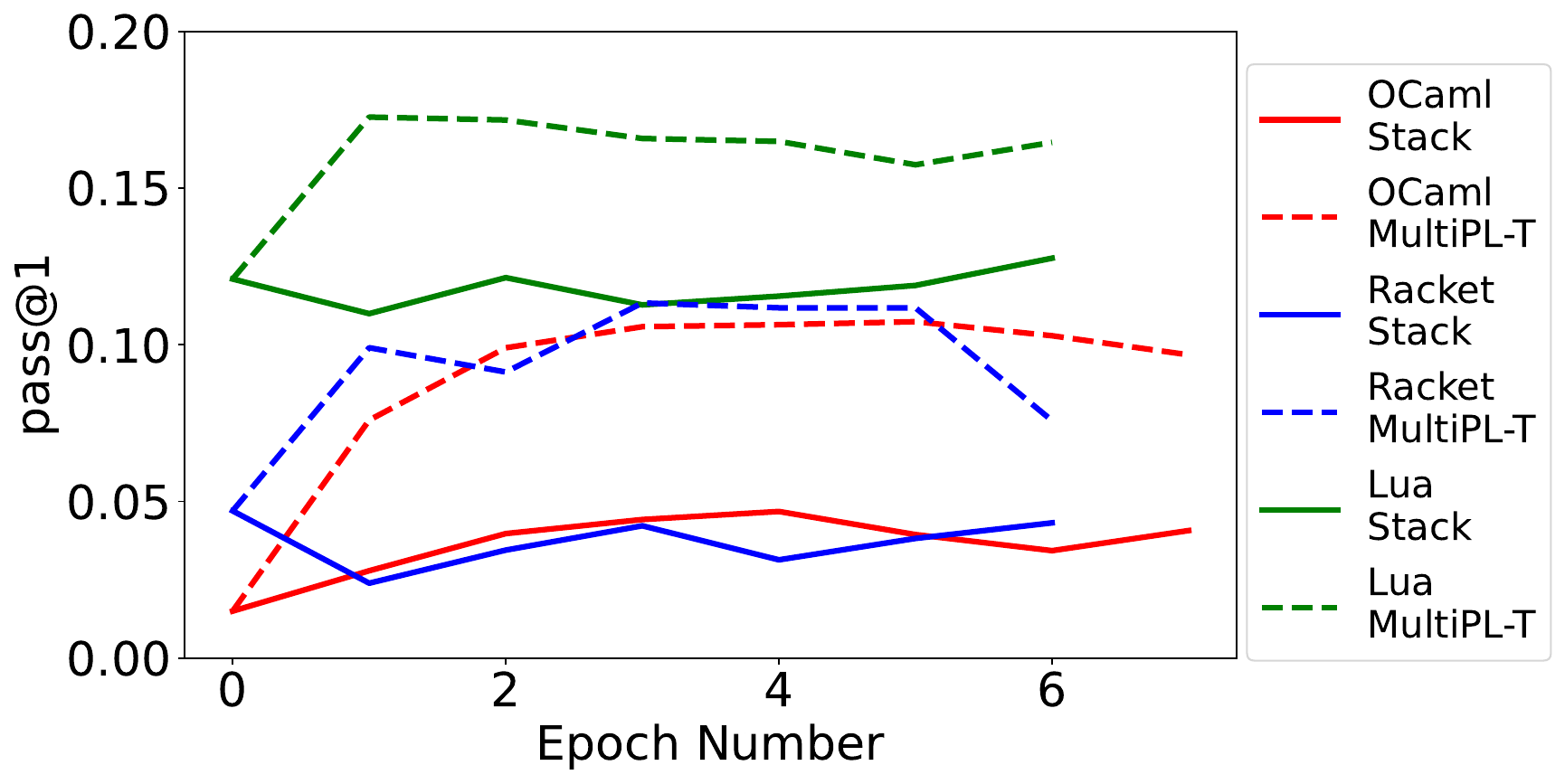}
\caption{Fine-tuning on subsets that are approximately the same size as the \system{} fine-tuning datasets.}
\label{strawman-just-train-more-subset}
\end{subfigure}

\caption{We fine-tune StarCoderBase-1B on several epochs of language-specific data of The Stack and measure performance with MultiPL-E. In \cref{strawman-just-train-more-full}, we train on all data from The Stack for each language. These datasets vary in size (the labels measure their size in tokens).
In \cref{strawman-just-train-more-subset} we sample each dataset to be approximately the same size as the MultiPL-T datasets. Both approaches that use data from The Stack barely improve performance, and can even hurt performance. 
In contrast, fine-tuning on \system{} (dashed lines) shows significant improvement.}
\label{strawman-just-train-more}
\Description{}
\end{figure}

\subsection{Further Training on Natural Data}
\label{just-train-more}
The simplest way to boost the performance of a Code LLM on a programming language is to train it further on \emph{natural data}, which is code written by human programmers rather than code generated by other means (e.g., an LLM). This was the approach taken to create StarCoder from StarCoderBase. The latter is the base model, and the former is fine-tuned on roughly two additional epochs\footnote{An \textit{epoch} in machine learning refers to one complete pass through the entire training dataset.}  of the Python subset of The Stack. Although this approach is effective for high-resource languages, we now show that it does not work for several low-resource languages (\Cref{strawman-just-train-more}).

In \cref{strawman-just-train-more-full}, we fine-tune three versions of StarCoderBase-1B on three more epochs of Lua, OCaml, and Racket each. This data is from The Stack. However, 
The Racket subset of The Stack is poor quality, so we use the Scheme subset instead.\footnote{The Racket subset accidentally omits the \texttt{.rkt} file extension and largely contains Racket documentation (in Scribble). Since Racket is descendant from Scheme, the Scheme subset is a more reasonable fine-tuning set.}
The Stack has an order of magnitude more Lua than OCaml and Racket. Moreover, even the Racket and OCaml data in The Stack is significantly larger than the fine-tuning datasets we will develop with MultiPL-T. Therefore, these experiments are not directly comparable to each other, since they train on wildly varying amounts of data. Nevertheless, we get poor results for all: the performance of these fine-tuned models barely increases for Racket and OCaml and even decreases for Lua.

In \cref{strawman-just-train-more-subset}, we do another experiment with The Stack that lends itself to a direct comparison with \system{}. We randomly sample data from The Stack to get approximately the same volume of data that we generate with \system{}. Thus, fine-tuning on these datasets will use similar computing resources as fine-tuning a model with \system{} data.
We use this data to fine-tune three versions on StarCoderBase-1B for six epochs, and evaluate the models at each epoch. We still get poor results with The Stack: Lua and OCaml performance barely increases and Racket performance decreases. In contrast, fine-tuning with \system{} will show significant gains.

\begin{figure}
\setlength{\belowcaptionskip}{10pt}

\begin{subfigure}{\textwidth}
\begin{lstlisting}[style=Racket]
;; sum-digits-of-number: Number -> Number
;; Sums the digits of a number
(define (sum-digits-of-number num)
  (foldr + 
         0 
         (map string->number ~\colorbox{pink}{(string-split (number->string num) "")}~)))  
\end{lstlisting}
\caption{Splitting on the empty string produces empty strings in the list, which cannot be converted to a number.}
\end{subfigure}

\begin{subfigure}{\textwidth}
%cspell:disable
\begin{lstlisting}[style=Racket]
;; add-rev-nums: [List-of Numbers] -> Number
;; Adds the reverse of all the numbers in a list
(define (add-rev-nums lon)
  (foldr + 
         0 
         (map (lambda (n) 
                (string->number (list->string 
                  (reverse (string->list (number->string n)))))) 
                    lon))) 
\end{lstlisting}
%cspell:enable
\caption{Working function.}
\end{subfigure}

\begin{subfigure}{\textwidth}
\begin{lstlisting}[style=Racket]
;; is-palindrome-sum?: Number -> Boolean
;; Checks if a number is a palindrome sum of its digits
(define (is-palindrome-sum? num)
  (is-palindrome? (sum-remainders 
    (map ~\colorbox{pink}{char->integer}~ (string->list (number->string num))) 10)))
\end{lstlisting}
\caption{\texttt{char->integer} returns the ASCII code of a character, and does not parse it as a digit.}
\end{subfigure}

\begin{subfigure}{\textwidth}
\begin{lstlisting}[style=Racket]
;; sum-palindromes: [List-of Numbers] -> Number
;; Sums the palindrome numbers in the list
(define (sum-palindromes lon)
  (foldr + 0 (filter ~\colorbox{pink}{is-palindrome?}~ lon)))
\end{lstlisting}
\caption{Utilizes what looks like a helper function, \texttt{is-palindrome?}, which is not defined.}
\end{subfigure}

\begin{subfigure}{\textwidth}
%cspell:disable
\begin{lstlisting}[style=Racket]
;; is-prime?: Number -> Boolean
;; Checks if a number is prime
(define (is-prime? num)
  (cond
    ((= num 2) #t)
    ((= num 1) #f)
    (else (zero? (remainder ~\colorbox{pink}{(add-odds (repeat "2" (/ num 2)))}~ num)))))
\end{lstlisting}
%cspell:enable
\caption{The highlighted code calls two helper functions that are not defined.}
\end{subfigure}

\caption{Faulty Racket code generated by StarCoderBase-15B when seeded with five hand-written examples.}
\label{racket-faults}
\Description{}
\end{figure}
\subsection{Self-Instruction for Low-Resource Programming Languages}
\label{use-self-instruct}

An alternative to fine-tuning on natural data is to fine-tune on LLM-generated data~\cite{wang:self-instruct,ziyang:wizard-coder}. The usual approach is to hand-select a seed dataset of programs, prompt the model with each seed to generate more programs, and iterate until a large enough dataset has been collected. This type of approach has been used successfully to generate training data for Code LLMs in high-resource languages~\cite{ziyang:wizard-coder}.
However, it should be obvious that \emph{these approaches presuppose that the LLM is good at generating reasonably correct and high-quality programs.} We are interested in languages that the model is bad at, so it should not be surprising that self-instruction does not work.

\paragraph{Illustrating self-instruction}
To illustrate how self-instruction goes wrong with low-resource languages, we use StarCoderBase-15B to generate functions in Racket, mimicking the first step of self-instruction. We prompt the model with five hand-written examples 
(\ifappendix \Cref{appendix:human-examples} \else included in the supplementary material\fi) and have it generate five more functions (\Cref{racket-faults}).
We find that four of the five model-generated programs have bugs.
This is a much higher error rate than what is evident from self-instruct datasets for high-resource languages~\cite{codealpaca,muennighoff:octopack}.
% NOTE(arjun): I asked Armel Randy, who says that the OctoPack paper is the preferred citation for the Self-Instruct StarCoder dataset.

\paragraph{A self-instruction experiment}
In \ifappendix \Cref{appendix:full-self-instruct} \else the supplementary material \fi we present results from an experiment where we self-instruct StarCoderBase-15B on Racket and get the expected poor results.

\vskip 1em
\noindent
Self-instruction and training further on existing public data do not help Code LLMs perform better on low-resource programming languages. Thus we now turn to the \system{} approach.

\begin{figure}[t]
\centering
\includegraphics[width=\textwidth]{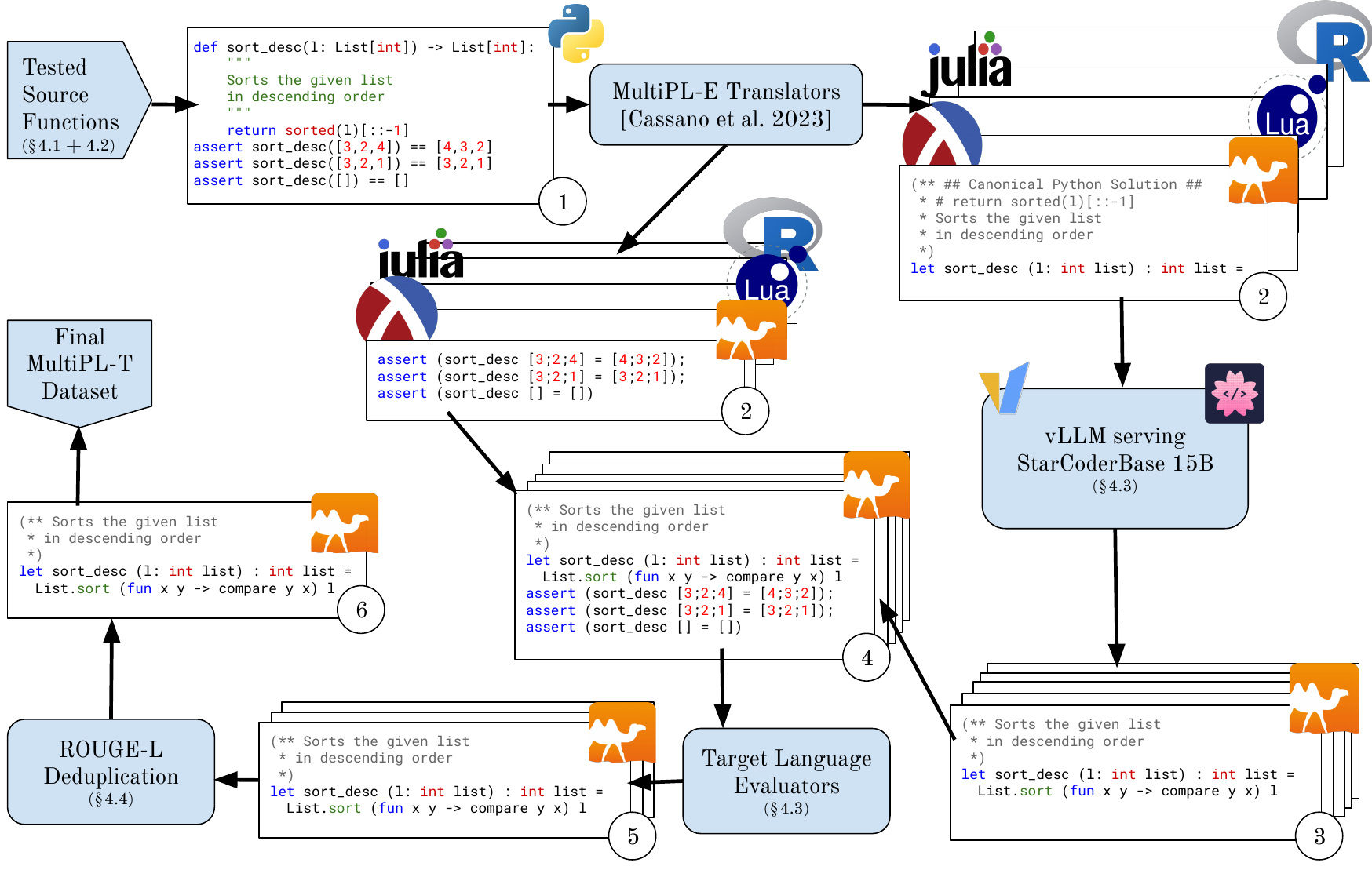}
\caption{The \system{} pipeline for generating semi-synthetic data. Starting with a tested Python function \circnum{1}, we compile the header and test cases into each target language \circnum{2}. We prompt the model with the target-language header and comment to 
generate  50-100 candidate translations \circnum{3} (varies by language). We append the compiled test cases to the candidate translations \circnum{4} and evaluate with the target language runtime. We deduplicate all programs that pass all test cases \circnum{5} to build the fine-tuning dataset \circnum{6}.}
\label{fig:codegenPipeline}
\Description{}
\end{figure}

% NOTE(arjun): The "Our Approach" section name is standard, and will help a reader trying to skip to what
% we actually do. Do not rename it to something more creative.
\section{Our Approach}
\label{sec:multiplt}

We now present the \system{} approach to generating high-quality, semi-synthetic data for low-resource languages.
\Cref{fig:codegenPipeline} depicts the \system{} system, which has several stages.
\begin{inparaenum}[\bfseries 1)]
\item Given a training dataset (The Stack), we filter data from a high-resource language (Python) to select code that is amenable to automatic test generation and translation. The Stack has 60GB of Python, and translation and test generation are expensive, so we filter quite aggressively. We only select individual Python functions that have docstrings and pass a heuristic type-checker (\cref{sec:py-source-funcs}).
\item Given the filtered dataset, we use a Code LLM (StarCoder-15B) to generate test suites for each function. We validate the generated tests for correctness and code coverage, and find that the Code LLM can be used as a capable test generator for our purposes (\cref{generating-unit-tests}). 
\item We translate each Python function to a target language $L$, by prompting the Code LLM to translate code. This translation may go wrong, especially because the Code LLM performs poorly on the low-resource target language.
\item We filter the $L$ functions (from Step 1) to only select those that pass test cases. To do so, we compile the Python test cases (from Step 2) to the language $L$, using the Python-to-$L$ test case compiler from MultiPL-E. The test case compiler is a traditional compiler that does not suffer from LLM hallucinations: if it cannot compile a test case, it signals an error, and we discard the training item if too many test cases fail to compile (\cref{sec:gen-target-funcs}).
\end{inparaenum}
The final result is thus a dataset of novel training items for the language $L$, which may be used to fine-tune any LLM. In \cref{evaluation}, we discuss how we use this data to fine-tune and evaluate several models for five different low-resource languages. The rest of this section describes the above steps in depth.

\subsection{Filtering Data from a High-Resource Language for Translation and Test Generation}
\label{sec:py-source-funcs}

\begin{table}
\caption{Size of the Python source dataset after each filtering step.}
\begin{tabular}{|l|r|}
\hline
     \textbf{Filtering Step} & \textbf{\#Functions}\\ 
     \hline
     All functions & 22,311,478\\
     With docstrings & 5,359,051 \\
     Typechecked and returns value & 459,280 \\
     No TODOs and no benchmark solution & 432,361\\
     Test generation & 157,767 \\
     90\% line coverage from tests & 133,168  \\
\hline
\end{tabular}
\label{fig:sourceFiltering}
\end{table}
% \begin{figure}[h]
%     \centering
%      f\includegraphics[width=0.75\textwidth]{./figures/source_filtering_stats.pdf}
%     \caption{Size of the Python source dataset after each filtering step.\aj{I will create a table.}}
%     \label{fig:sourceFiltering}
% \end{figure}

The first step in \system{} is to filter code from a high-resource language to serve as the translation source for our semi-synthetic data. We use Python because it has the highest representation in The Stack and because MultiPL-E can compile Python function signatures and test cases to several low-resource languages. However, our approach could easily be adapted to work with other high-resource languages.

\paragraph{Filtering Python Functions Before Translation}

The Stack has 22 million Python functions (\Cref{fig:sourceFiltering}). However, not all of these are amenable to translation and test-based validation with \system.
One could naively try to translate and generate tests for all 22M functions. However, since doing so requires GPUs, it would be prohibitively expensive.
Instead, we aggressively filter the 22M functions down to \textasciitilde400,000 functions using the following steps:
\begin{enumerate}

\item We exclude Python functions that do not have a docstring or use non-ASCII characters. One could generalize to include functions that have an associated comment. However, we still end up with over 5M candidate functions with this simple filter.

\item We use the Pyright~\cite{pyright:github} Python checker to validate that each function returns a value, uses only the Python standard library, and is thus likely type-correct. Pyright uses heuristics and makes no attempt at being sound. This does not impact \system{}, since we merely use typeability as a heuristic for code quality. This narrows the 5M functions to approximately 460,000 functions.

\item We exclude Python functions that have comments suggesting the implementation is incomplete (e.g. ``TODO''). It turns out that a fair amount of code on The Stack is incomplete; these functions are not likely to be useful training data. To avoid data contamination, we filter out functions whose prompt or solution appears in widely-used Code LLM benchmarks by finding exact matches of the prompts~\cite{chen2021evaluating,austin2021program}.

\end{enumerate}

The final dataset contains 432,361 Python functions. With this narrower set of functions, we move on to the next steps that require GPUs.

\begin{figure}
    \centering
    \begin{subfigure}{.45\textwidth}
        \centering
        \includegraphics[width=1\textwidth]{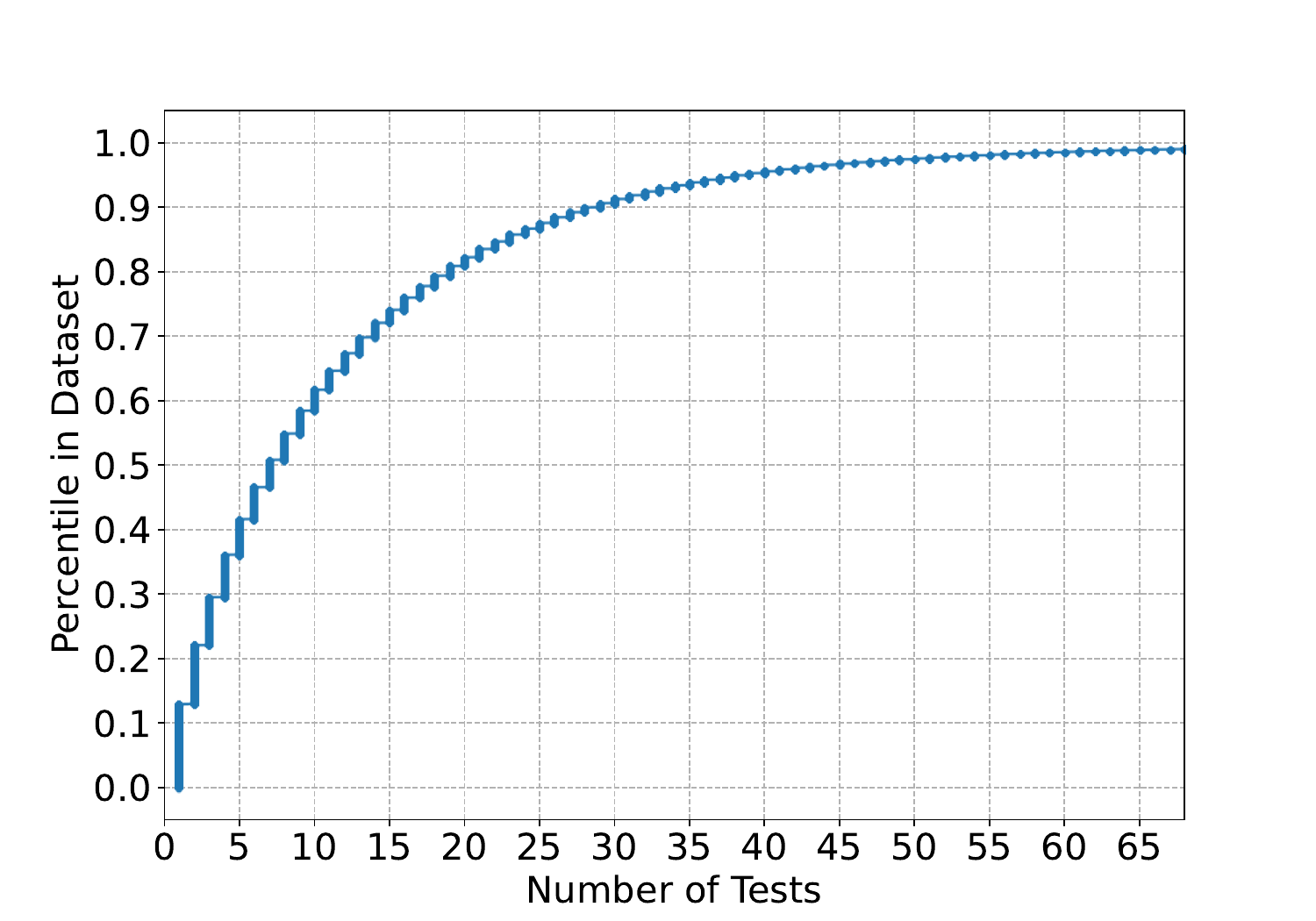}
        \caption{The $y$-axis shows the fraction of test suites with fewer than $x$ tests.}
        \label{fig:promptTranslation:testSizes}
    \end{subfigure}
    \hfill
    \begin{subfigure}{.45\textwidth}
        \centering
        \includegraphics[width=1\textwidth]{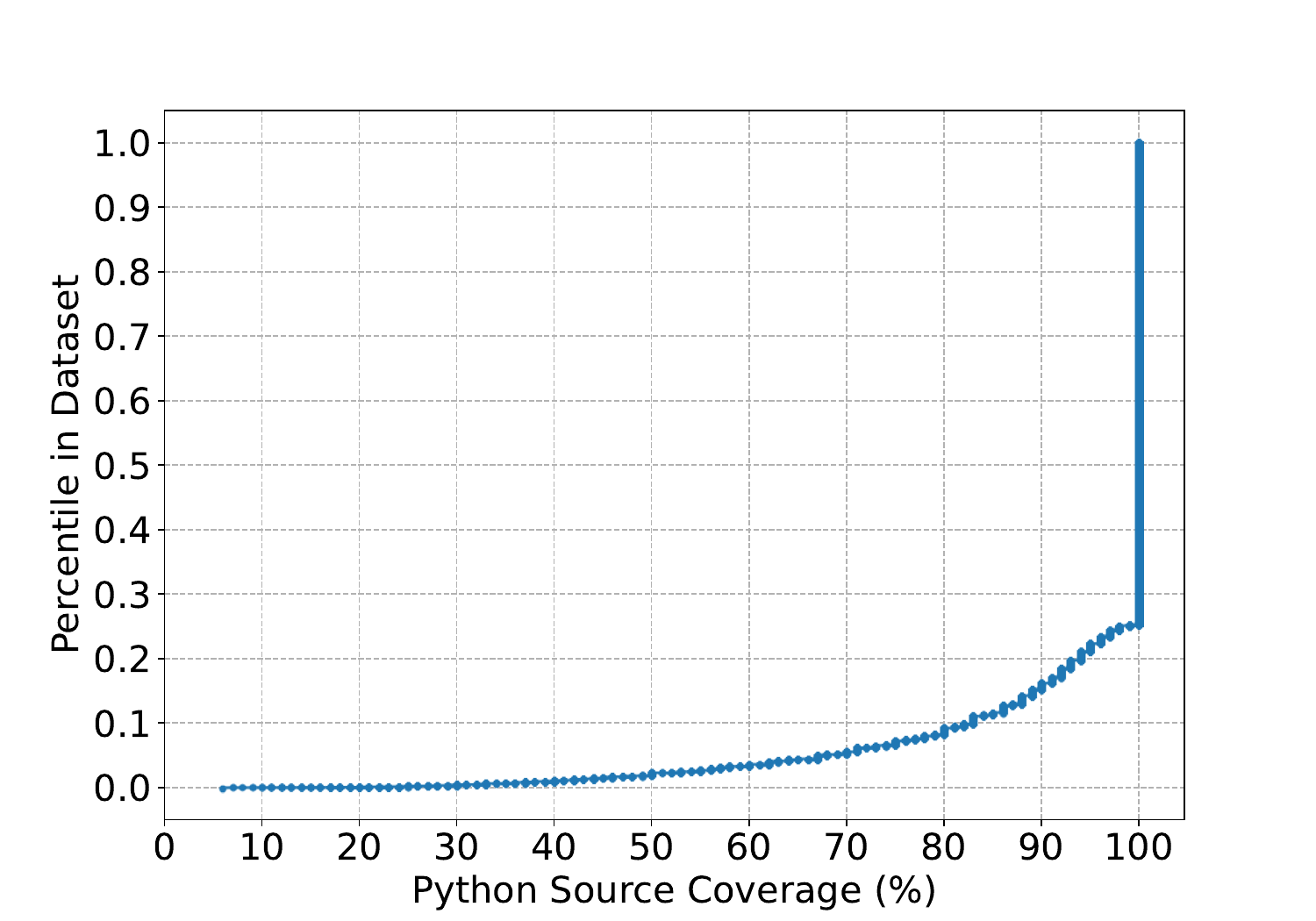}
        \caption{The $y$-axis shows the fraction of test suites with less than $x$\% coverage.}
        \label{fig:promptTranslation:coverage}
    \end{subfigure}
    \caption{Python test suite sizes and coverage distribution among the unfiltered test generation dataset.}
    \label{fig:promptTranslation}
\Description{}
\end{figure}

\begin{figure}
\begin{lstlisting}[style=codeblock, language=Python]
def count(word):
    """Count the number of X's in a word."""
    count = 0
    for letter in word:
        if letter == "Y":
            count += 1
    return count

~\colorbox{pink}{assert count("XXXXXX") == 6}~ # First completion
assert count("YYY") == 3 # Second completion
\end{lstlisting}
\caption{An example prompt to generate a unit test where the code and comment are inconsistent: the comment counts X's, but the code counts Y's. Using a Code LLM to generate tests helps expose the inconsistency: we get some tests that are based on the comment and others based on the code. We show two completions we got from StarCoderBase-15B in two successive queries. The ``\texttt{assert count(}'' is part of the prompt, but we show it with the completion for clarity.}
\label{inconsistent-training-example}
\Description{}
\end{figure}

\subsection{Generating Python Unit Tests}
\label{generating-unit-tests}

The next step in \system{} is to generate unit tests for each Python function. We will then compile these unit tests to target low-resource languages using the test case translators from MultiPL-E. We generate Python unit tests using the following steps.

\begin{enumerate}

\item We prompt the Code LLM to generate assertions given the Python function, taking care to generate several independent test suites to get high coverage, and executing the tests to ensure that they pass.

\item For each function, we measure the coverage of the aggregated test suites, discarding functions with less than 90\% line coverage.

\item We use the tests to infer basic types for the Python function, which is necessary to compute the tests to certain target languages.

\end{enumerate}

\paragraph{Test generation}
Instead of using a traditional test generator that synthesizes tests from code (e.g., \cite{pynguin}), we use a Code LLM to generate test cases by prompting the model to produce an assertion. The Code LLM conditions on the source code text and thus serves as a weak detector for inconsistencies between code and comments.
For example, \cref{inconsistent-training-example} shows a function that would be a bad training item: the code counts Y's, but has a comment that says it counts X's. When we generate several test cases independently, we end up with tests for both X and Y, which lowers the ratio of passing tests. Conditioning on the text makes it less likely that inconsistent tests will be generated, decreasing the likelihood that functions will be filtered out of the training set due to low coverage (as described below).

We prompt StarCoder-15B to generate five independent test suites for each function with high temperature\footnote{\textit{Temperature} is a parameter that controls the randomness of the next-token predictions. A higher temperature results in more varied (less predictable) output, while a lower temperature produces more conservative (more predictable) results.} (0.8) to get a diverse set of candidate tests. We parse each generated test suite and extract all test cases that are suitable for translation using MultiPL-E. We take the set of matching test cases and run each test in isolation in a container to verify that it passes, discarding any that fail. If no correct tests are generated, we discard the function. The result is a set of nearly 160,000 Python functions with at least one passing test case.
\Cref{fig:promptTranslation:testSizes} shows the distribution of test suite sizes. The median number of test cases per function is 7, and the mean is 12.1. Note that we do not filter any tests after validated, so a function may be more tests that strictly necessary.

\paragraph{Filtering on test coverage}

Given the dataset of Python functions with docstrings and test suites, our next step is to filter out functions with low test coverage.
We use \emph{line coverage} as the coverage metric and exclude all functions with less than 90\% line coverage. The result is a dataset of 133,668 functions with 90\% line coverage from tests.

Since we start with nearly 160,000 functions, this implies that most of the generated test suites that work have high line coverage. In fact, most functions have 100\% line coverage (\Cref{fig:promptTranslation:coverage}).
This stringent criterion ensures that the functions in our final set are not just correct but also comprehensively tested, reinforcing the reliability of our dataset. In the filtered Python dataset, the average function has 10.3 lines of code (SD=10.7) and on average 3.6 branches (SD=4.1).

\begin{figure}

\begin{subfigure}{0.45\textwidth}
\begin{lstlisting}[style=codeblock, language=Python]
def rep_or_hello(x):
    """Repeat string twice, or
    produce Hello if no string given"""
    if x is None:
        return "Hello"
    else:
        return x + x

assert rep_or_hello(None) == "Hello"
assert rep_or_hello("y") = "yy"
\end{lstlisting}
\caption{Python function and generated tests.}
\label{ocaml-ex-lhs}
\end{subfigure}
\quad
\begin{subfigure}[b]{0.45\textwidth}
\begin{lstlisting}[style=codeblock, language=ML]
(* Repeat string twice, or produce
   Hello if no string given. *)
let rep_or_hello (x : string option) 
  : string = 

assert rep_or_hello None == "Hello";;
assert rep_or_hello (Some "y") = "yy"
\end{lstlisting}
\caption{OCaml prompt and compiled tests.}
\label{ocaml-ex-rhs}
\end{subfigure}
\caption{An example that demonstrates the need for Python type inference when translating to OCaml. \Cref{ocaml-ex-lhs} shows a Python function with generated test cases. We then compile both the Python function signature and the test cases to OCaml (\cref{ocaml-ex-rhs}). However, to know that the Python string \texttt{"y"} must compile to \texttt{Some "y"} in OCaml, we need to infer Python types.}
\label{ocaml-ex}
\Description{}
\end{figure}

\paragraph{Type inference}

The steps described above are sufficient to generate data for an untyped low-resource language (e.g., Racket or Lua).
However, for a typed target (e.g., OCaml or Julia), we also need to infer types for two reasons.\footnote{Julia can be used with or without types, and our dataset has both typed and untyped examples.} Consider the case where we target OCaml. First, we rely on the LLM to generate an OCaml function body, given only a comment and the function header (\texttt{let f x =}). Without type annotations, the only signal about the desired type of \lstinline|f| are the identifier names and the comment. With type annotations, the LLM is far more likely to produce a function with the expected type.  Second, we need to infer Python types to compile test cases, which we illustrate in \cref{ocaml-ex}. In this example, we have a Python function that consumes an optional string. We have two test cases, one that applies the function to \texttt{None} and the other that applies the function to a string. When we compile these tests to OCaml, we have to transform the Python argument type \texttt{Union[str,None]} to the OCaml type \texttt{string option}. We can only do this type transformation after we have inferred the Python type.

Our approach to type inference is simple: we deduce types based on test cases, ignoring the function body. We extract the instance type of each argument and expected return value in each test, computing the union type between the types at the same position among tests. 
For example, if the test cases apply \texttt{foo(1)} and \texttt{foo(None)}, we infer \texttt{Union[int, None]} as the type of \texttt{foo}'s argument. Moreover, we simplify \texttt{Union[T, None]} to the more canonical \texttt{Optional[T]}. For example, \texttt{Union[int, int, None]} would be then simplified to \texttt{Optional[int]}.
This approach to type inference can only fail if the code is non-deterministic, which does not occur in our dataset.

\vskip 1em
\noindent
Following the steps above produces two datasets of Python functions---one with and one without type annotations---where every function has a docstring and a suite of unit tests that achieve high coverage. These datasets can be used to generate training data for any low-resource language.

\subsection{Translation from a High-Resource to a Low-Resource Language}\label{sec:gen-target-funcs}

Given a dataset of commented Python functions with high coverage unit test suites, our next goal is to translate the dataset from Python to a target language $L$ and use tests to validate the translation.

\paragraph{Translation with a Code LLM and MultiPL-E}

We use a modified version of MultiPL-E~\cite{cassano:multipl-e} to translate each Python function into an equivalent function in the target low-resource language. 
We construct a MultiPL-E prompt with the following three parts:
\begin{enumerate}
    \item \emph{Docstring:} We turn the Python docstring  into a comment in the target language. The MultiPL-E toolchain translates between different comment formats and also alters common type names in natural language using simple rules. For example, when translating from Python to OCaml, we turn ``dictionary'' into ``association list''.
    
    \item \emph{Function signature:} We turn the Python function signature into a function signature in the target language. This step may involve translating types from Python into the target language if they are required.

    % NOTE(arjun): Molly had asked, "the Figure suggests it is always Python code, then translated docstring. Why this order?". The answer is that we hadn't tried the other order. It is just another variation in the infinite space of ablations and very unlikely to matter much. My intuition is that it would be worse. [Molly thanks you for the answer!]
    \item \emph{Original Python code:} Finally, we add a comment (in the target language) that contains the original Python code. We find that this additional information increases the chance that the model generates a correct translation (\Cref{sec:prompt-ablation}).

\end{enumerate}

\Cref{fig:codegenPipeline} highlights an example prompt and test suite for translating a descending sort function written in Python to OCaml in the programs labeled 1 and 2.
MultiPL-E translates comments written in Python to OCaml and translates each test case and the function signature from Python to OCaml. The original Python code is added as part of the comment.

Given this prompt, we use StarCoderBase-15B to generate translations of each problem in our Python dataset. For all of our languages, we generate 50 translations\footnote{For OCaml, we generated 100 translations per problem, as the base pass rate was significantly lower than other languages.} with high temperature (0.8), to encourage the Code LLM to produce a more diverse set of candidate solutions~\cite{chen2021evaluating}.

% Commenting out this figure because it's data is redundant with fig 5
% \begin{figure}[t]
%     \centering
%     \begin{subfigure}{.45\textwidth}
%         \centering
%         \begin{lstlisting}
% def sort_desc(l: List[int]) -> List[int]:
%    """
%    Sorts the given list in
%    descending order
%    """
%    return sorted(l)[::-1]
%         \end{lstlisting}
%         \caption{Python function}
%         \begin{lstlisting}
% assert sort_desc([3, 2, 4]) == [4, 3, 2]
% assert sort_desc([3, 2, 1]) == [3, 2, 1]
% assert sort_desc([]) == []
%         \end{lstlisting}
%         \caption{Python test suite}
%     \end{subfigure}
%     \hfill
%     \begin{subfigure}{.45\textwidth}
%         \centering
%         \begin{lstlisting}
% (** ## Canonical Python Solution: ##
%  * # return sorted(l)[::-1]
%  * Sorts the given list in
%  * descending order
%  *)
% let sort_desc (l: int list) : int list =
%         \end{lstlisting}
%         \caption{Translated OCaml prompt}
%         \begin{lstlisting}
% let () =
% assert (sort_desc [3; 2; 4] = [4; 3; 2]);
% assert (sort_desc [3; 2; 1] = [3; 2; 1]);
% assert (sort_desc [] = []);
%         \end{lstlisting}
%         \caption{Translated OCaml test suite}
%     \end{subfigure}
%     \caption{Example of prompt and tests translation for a descending sort function from Python to OCaml.}
%     \label{fig:promptTranslation}
% \end{figure}

\paragraph{Checking translations with compiled tests}

A Code LLM is quite likely to produce faulty translations; in our case, this is even more likely, since we specifically target languages on which the Code LLM performs poorly. We address this problem by translating test cases from Python to the target language using a simple, recursive compiler.
MultiPL-E has a suite of compilers that translate simple Python assertions into assertions in 20+ other programming languages.
The compilers support assertions that are simple equalities between first-order values, specifically atomic Python data and collections (lists, tuples, and dictionaries).
We use these compilers to translate tests from Python to each target language, removing test cases that MultiPL-E does not support. If we are left with zero test cases, we discard the function entirely.

Given the set of 50 generated translations for each function, we select only those solutions that pass all tests. This may include selecting several solutions to the same problem, which is beneficial to the model in terms of learning diverse code styles.

\subsection{Deduplication}
\label{sec:dedup}

\begin{algorithm}[t]
\caption{Parallelized ROUGE-L deduplication procedure.}
\label{dedup-algo}
\begin{algorithmic}[1]
\Procedure{Deduplicate}{$I$, $t$, $\mathit{groupSize}$, $\mathit{rounds}$} \Comment{Deduplicate items ($I$) with threshold $t$}
  \State $G \gets$ \Call{GroupByPrompt}{$I$} \Comment{Group items by prompt (comment)}
  \State $\mathit{deduped} \gets$ \Call{DedupGroups}{$G$,$t$} \Comment{Deduplicate for each prompt}
  \For{$i \gets 0$ to $rounds$}
    \State $G \gets$ \Call{RandomGroups}{$\mathit{deduped}$, $\mathit{groupSize}$} \Comment{Randomly group items}
    \State $\mathit{deduped} \gets$ \Call{DedupGroups}{$G$, $t$} \Comment{Global deduplication}
  \EndFor
  \State \Return $\mathit{deduped}$
\EndProcedure

\Procedure{DedupGroups}{$G$, $t$}
\Comment{Deduplicates item within each group in the list of groups ($G$).}
  \State $\ell \gets []$
  \For{$g \gets G$ \textbf{in parallel}}
    \State $\mathit{keep} \gets [ \textbf{true} \mid x \in g ]$ \Comment{Initially keep every item in the group}
    \For{$i \gets 0$ to $\mid g \mid - 1$}
        \For{$j \gets i+1$ to $\mid g \mid$}
            \If{$i = j$ or not $\mathit{keep}[j]$}
                \State \textbf{continue}
            \EndIf
            \State $a, b \gets $\Call{RemComments}{$g[i]$}, \Call{RemComments}{$g[j]$}
            \If{$\text{F-Measure}(\text{ROUGE-L}(a,b))  > t$}
                \State $\mathit{keep}[j] \gets \textbf{false}$ \Comment{Remove an item if it is similar to others in the group.}
            \EndIf
        \EndFor
    \EndFor
    \State $\ell \gets \ell +  [ g[i] \mid \mathit{keep}[i] ]$
  \EndFor
  \State \Return $\ell$ \Comment{A list of items (ungrouped)}
\EndProcedure

\end{algorithmic}
\end{algorithm}

We define a set of solutions as \textit{diverse} when they differ in form.
For instance, two solutions that perform identically on a test suite but differ in their implementation--one using a recursive function and the other a loop--are considered diverse.
Ensuring diversity among generated and verified solutions is crucial for teaching fine-tuned models a variety of syntactic and semantic features of the target programming language. Furthermore, redundant or similar solutions may diminish the effectiveness of a dataset~\cite{lee2022deduplicating}.

Just resampling with high temperature does not guarantee diverse solutions: the LLM may still produce nearly identical solutions (e.g., with a few variables renamed). To address this, we employ a deduplication algorithm based on  ROUGE-L~\cite{lin-och-2004-automatic}.
ROUGE-L is a metric of text summarization quality, and quantifies the syntactic overlap between two pieces of text with a score ranging between $0$ and $1$ where $1$ indicates the highest similarity. We use 0.6 as the similarity threshold for discarding duplicates. Before comparing a pair of solutions, we remove all comments from the code, as it may introduce noise in the deduplication process.

Running ROUGE-L on all pairs of items is prohibitively expensive. Instead, we use a heuristic that is amenable to parallelization (\Cref{dedup-algo}). We apply ROUGE-L to deduplication items in fixed-size groups (we use size 200). Initially, the groups are solutions to the same prompt, as this group is likely to have many duplicates. We then randomly regroup items and run grouped deduplication again. The number of rounds of deduplication is proportional to the total number of items: more rounds increase the likelihood that duplicates will be removed. Ultimately, this results in a set of diverse, accurate, and semantically equivalent solutions for each prompt.

% People expect this section to called "Evaluation" and not something more creative.
\section{Evaluation}
\label{evaluation}

In this section, we use \system{}  to fine-tune a variety of different models of varying sizes. We demonstrate state-of-the-art results on standard benchmarks for the natural language to code task (\cref{standard-eval}), novel qualitative and quantitative evaluation (\cref{new-eval}), and ablations showing the significance of various design decisions in \system (\cref{ablations}).

\subsection{Experimental Setup and Implementation}

\paragraph{Training hyperparameters}

We fine-tune all models with a sequence length of 2,048 tokens. StarCoderBase-1B is fine-tuned for seven epochs with batch size $8$, with learning rate $3 \times 10^{-5}$, $10$ steps warmup, and cosine learning rate decay. For StarCoderBase-15B, CodeLlama-34B, and CodeLlama-70B we make these configuration changes: ten epochs, batch size $32$, and learning rate $2 \times 10^{-5}$.

\paragraph{Estimated computing resources used}

The work for this article was done over several months using V100 (32GB), A100 (80GB), and H100 (80GB) NVIDIA GPUs, as they were available across several clusters and servers. We estimate that we spent approximately 550 days of A100 (80GB) GPU time with the following breakdown:
\begin{itemize}
  \item Training: \textasciitilde3,444 hours fine-tuning several versions of CodeLlama-70B, CodeLlama-34B, StarCoderBase-15B, and StarCoderBase-1B. These include the models presented in this section and in \cref{alternatives}.
  \item Evaluation: \textasciitilde310 hours running benchmarks, which include MultiPL-E and the new in-context learning benchmark (\cref{sec:in-context}).
  \item Dataset Generation: \textasciitilde9,984 hours generating the \system{} training sets. This was the most significant use of resources but is reusable for future model development.
    Using models larger than StarCoderBase-15B for generation would take up significantly more resources.
\end{itemize}

\paragraph{Training and evaluation tools}
We experimented with several technologies during development. The final \system{} pipeline uses vLLM~\cite{vllm} for inference, DeepSpeed ZeRO to fine-tune larger models~\cite{deepspeed-zero}, Transformers~\cite{hf-transformers}, and MultiPL-E~\cite{cassano:multipl-e} with several modifications, such as supporting OCaml.

\subsection{Evaluation on Standard Benchmarks}
\label{standard-eval}

Most Code LLM benchmarks target high resource programming languages, mostly Python. To evaluate our fine-tuned LLMs for low-resource languages, we use and extend MultiPL-E, which is the benchmark that was used to evaluate StarCoder, Code Llama, Stability.ai's StableCode, and several other Code LLMs. As described in \cref{about-multipl-e}, MultiPL-E is a benchmark for the natural language to code task. We report MultiPL-E results in the standard way: using the pass@1 metric, which is simply the \emph{mean pass rate} on the benchmark, where a passing completion must successfully execute all tests.

\paragraph{Choosing baselines and comparisons}

Before presenting results, we discuss how we believe fine-tuned models should be evaluated.

It is well known that larger models perform better because they can learn more complex patterns~\cite{hoffmann2022an}. Moreover, there is a recent trend of building smaller models that outperform larger models by training them on far more data~\cite{harms_law}. 
For example, StarCoder2-15B is the same size as StarCoderBase-15B, but is trained on 400\% more data and performs 50\% better on MultiPL-E~\cite{starcoder2}.

Therefore, it trivial to achieve an impressive fine-tuning result by simply starting with a newer, better model. We argue that the right way to evaluate a fine-tuning approach is to do all of the following:
\begin{enumerate}
    \item Compare the new fine-tuned model to the base model. This allows us to ask, \emph{how does a  fine-tuning approach improve baseline performance for a particular model?}
    \item Compare the new fine-tuned model to other fine-tunes of the same model. Fine-tuning on a large dataset is likely to do better than fine-tuning on a smaller dataset. But, this allows us to ask, \emph{How efficient is one fine-tuning approach compared to another?}
    \item Fine-tune a variety of base models. This allows us to ask, \emph{does the fine-tuning approach generalize to different model sizes and families?} 
\end{enumerate}
Thus a fine-tuning approach should improve baseline performance on several different model families and across several model sizes, and we will show that this holds for \system.

\begin{table}
\caption{MultiPL-E pass@1 scores for 1B, 15B, 34B, and 70B parameter models before and after fine-tuned on \system{} data. The fine-tuned models show significant improvement. 
% \Cref{higher-pass-k} reports pass@5 and pass@10 results.
}  
\label{tbl:main-results}  
\begin{tabular}{|l|r|r|r|r|r|r|r|r|}
\hline
% NOTE(arjun): The model is called StarCoderBase. Do NOT change to
% StarCoder because that is a different model.
\multirow{2}*{Language}   & \multicolumn{2}{c|}{StarCoderBase-1B} & \multicolumn{2}{c|}{StarCoderBase-15B} & \multicolumn{2}{c|}{CodeLlama-34B} & \multicolumn{2}{c|}{CodeLlama-70B} \\
            \cline{2-3}                           \cline{4-5}                            \cline{6-7}   \cline{8-9}                
           &  Base & Fine-tuned         & Base & Fine-tuned        & Base & Fine-tuned & Base & Fine-tuned \\
\hline
OCaml                 & 1.5   & 9.7        & 6.9  & 19.9      & 18.3 & 27.4 & 23.2 & 29.3 \\
Racket                & 4.7   & 11.3       & 11.8 & 21.0      & 15.9 & 29.1 & 21.9 & 33.1 \\
R                     & 5.4   & 8.9        & 10.2 & 17.3      & 18.2 & 25.5 & 23.0 & 28.5\\
Julia                 & 11.3  & 15.6       & 21.1 & 35.2      & 31.8 & 43.5 & 41.9 & 44.5 \\
Lua                   & 12.1  & 17.3       & 26.6 & 31.0      & 38.1 & 43.9 & 41.7 & 44.9 \\
\hline
\end{tabular}
\end{table}

\subsubsection{Fine-Tuning StarCoderBase and Code Llama}

Fine-tuning StarCoderBase and Code Llama with \system{}-generated data on our target languages improves performance on MultiPL-E across the board
(\Cref{tbl:main-results}).
For each model, we fine-tune a separate model for \alllangs.
We checkpoint and evaluate the models at each epoch and report the peak performance.
It is \emph{not} a goal of this paper to maximize MultiPL-E scores. In fact, the next section suggests that it would be easy to improve some of these scores by either training longer on the data we already have or by letting \system{} generate more data.
Beyond the generally improved performance, we draw several other conclusions:
\begin{enumerate}

\item Racket and OCaml, which are the lowest-resource languages that we evaluate, show the largest relative gains. For example, the fine-tuned versions of both models have more than double the score of their base models, with particularly large gains for OCaml.
  Even the fine-tuned 1B models perform comparably to the base 15B models at producing correct solutions in these languages.

\item Lua obtains relative gains of 42\% for 1B and 17\% for 15B. However, these gains are significant and put the fine-tuned models' Lua performance on par with the base models' performance on the highest-resource languages. For example, StarCoderBase-15B achieves $30.6$ on MultiPL-Python, and our fine-tuned model achieves $31.0$ on MultiPL-Lua.

\item Julia also shows a significant relative gain of 67\% for 15B, achieving a score on MultiPL-Lua that exceeds the base model's MultiPL-Python scores. 

\item We also fine-tune and evaluate CodeLlama-34B and CodeLlama-70B, where we see significant improvements as well.

\item In contrast to distillation, where a larger model produces training data for a small model, the \system{} approach allows a smaller model to improve the performance of much larger models. Specifically, we use data generated by StarCoderBase-15B and validated using the \system{} approach to improve the performance of models that are nearly 5x larger.

\end{enumerate}
%
%Overall, \emph{when compared to open models that are not fine-tuned on proprietary data, these fine-tuned models achieve state-of-the-art benchmark scores on \alllangs{} at 1B, 15B, 34B, and 70B model sizes.}

\ifappendix\Cref{mbpp-eval}\else The supplementary material \fi includes an evaluation with the MBPP benchmark. On MBPP, we show even larger relative gains with \system.

\begin{table}
\caption{MultiPL-E pass@1 scores for two recently released models on Racket. The fine-tuned models are significantly better, even though the \system{} datasets are from an older model.}
\label{tbl:new-models-results}  
\begin{tabular}{|l|r|r|r|r|}
\hline
\multirow{2}*{Language}   & \multicolumn{2}{c|}{StarCoder2-15B} & \multicolumn{2}{c|}{DeepSeekCoder-Base-33B} \\
            \cline{2-3}                           \cline{4-5} 
           &  Base & Fine-tuned         & Base & Fine-tuned \\
\hline
Racket     &  22.41   & 29.7       &  	23.3 & 36.6   \\
\hline
\end{tabular}
\end{table}

\subsubsection{Fine-Tuning StarCoder2 and DeepSeek Coder}

When this work was done, Code Llama was the best-performing base Code LLM with open model weights. However, two new base models were recently released: DeepSeek Coder~\cite{guo:deepseek} and StarCoder2~\cite{starcoder2}. To show that  \system{} remains relevant, we fine-tune DeepSeekCoder-33B and StarCoder2-15B on our Racket dataset. The aforementioned models are the two top-performing base models of their size at the time this article was written. In both cases, we find that the \system{} approach continues to work. Both models show significant improvement in their Racket capabilities (\cref{tbl:new-models-results}).

The datasets presented in this article are part of the StarCoder2 pretraining corpus as one of several sources of ``(Leandro's) High Quality'' data. All we can conclude is that they did not hurt performance on the target languages and that further fine-tuning still help (contrast with \cref{just-train-more}).

\subsubsection{Summary}

Using \system{}, we have fine-tuned models from four different model families and a wide range of model sizes (1B to 70B parameters). We find that our fine-tuned models outperform base models across the board, with the most significant gains for the lowest-resource languages and small to mid sized models.

The largest models (70B) do not show as much as the smallest models. However, our datasets are generated with a much smaller model (15B), and it is unusual that we can use a smaller model to improve the performance of a much larger model~\citep{burns2023weaktostrong}.

At the time of writing, our fine-tuned version of DeepSeek Coder outperforms every fine-tuned model on the BigCode Models Leaderboard for Racket~\cite{allal:bigcode-leaderboard}. The leaderboard models fine-tuned on an order of magnitude more data than ours, as well as models that distill proprietary models such as GPT-4. Thus we conclude that for low-resource languages, \system{} is significantly more data efficient than alternative approaches to fine-tuning.

% \ag{Do we need the follow paragraph given my reframing?}
% In our opinion, many of the fine-tuned models are not of scientific interest for a number of reasons.
% \begin{inparaenum}[1)]
%     \item Some models disclose no details about their training process. For all we know, they may have directly trained on benchmarks or performed self-instruction on benchmarks.
%     \item Other models are trained on an arbitrary mixture of data sources in an attempt to beat benchmarks, with no attempt to determine what leads to improved performance.
%     \item Finally, many models distill data from massive proprietary models, breaking their terms of use and making it hard to establish baselines. (E.g., consider training a small model directly on a proprietary dataset instead of distilling it from an API.)
% \end{inparaenum}

\begin{table}
\caption{The pass@1 scores of the original models and fine-tuned models on our new  benchmark that uses user-defined types, higher-order functions, helper functions, and non-standard libraries. The scores suggest that the 15B model may have overfit when fine-tuned on OCaml. However, the other models do the same or better after fine-tuning.}
\label{tbl:contextbench-results}  
\begin{tabular}{|l|r|r|r|r|}
\hline
% NOTE(arjun): The model is called StarCoderBase. Do NOT change to
% StarCoder because that is a different model.
\multirow{2}*{Language}   & \multicolumn{2}{c|}{StarCoderBase-1B} & \multicolumn{2}{c|}{StarCoderBase-15B} \\
            \cline{2-3}                           \cline{4-5}                   
           &  Base & Fine-tuned & Base & Fine-tuned \\
\hline
OCaml                 & 33.0   & 33.7      &    50.6  & 42.9 \\
Racket                & 19.3   & 22.7      &    28.4  & 41.3 \\
Lua                   & 26.3   & 46.9      &    48.7  & 51.3 \\
\hline
\end{tabular}
\end{table}

\subsection{New Tasks and Qualitative Evaluation}
\label{new-eval}

The previous section uses MultiPL-E for evaluation, which has a very particular format (\cref{about-multipl-e}) and uses unit tests to test correctness. Code LLMs are much more flexible and there are acceptability criteria that cannot be captured with unit tests or specific formats. We address these in this section.

\subsubsection{Evaluating In-Context Learning}
\label{sec:in-context}
A limitation of the \system{} datasets is that every training item is a single function without any other context: they may use standard libraries, but cannot depend on other functions, classes, or third-party libraries. Thus it is plausible that fine-tuning a model on \system{} data will make it overfit to this format.\footnote{Recall that the base models have been pretrained on natural code that is not constrained to the \system{} format. Thus appropriate fine-tuning should not eliminate their ability to generate code that is not in the \system{} format.}
Moreover, conventional Code LLM benchmarks, including MultiPL-E, will not expose this problem, because the benchmark tasks largely involve generating standalone functions that use only standard libraries.

To determine if this kind of overfitting is a problem, we construct a new, multi-language benchmark with fourteen problems.
We decided to manually construct each problem as opposed to sourcing them from repositories to ensure that the problems were similar across languages,
and more importantly, that they were not part of the training data.
Every problem has hidden test cases and a prompt that exercises the model's ability to use user-defined types, higher-order functions,  helper functions, or external libraries (\Cref{tbl:contextbench-desc}).  We manually translate these prompts into idiomatic OCaml, Racket, and Lua.

We evaluate StarCoderBase 1B and 15B on this new benchmark (\Cref{tbl:contextbench-results}). The results suggest that \emph{the 15B OCaml-tuned model may have overfit to the \system{} format.} However, \emph{all other fine-tuned models do the same or better}. 
We speculate that fine-tuning on a mix of natural and \system{} data will decrease the likelihood of overfitting.

\begin{table}
\caption{The 14  problems that exercise in-context learning.}
\label{tbl:contextbench-desc}

  \centering
  \begin{tabular}{|p{1.5in}|p{3.5in}|}
    \hline
    \textbf{Problem} &
    \textbf{Description} \\
    \hline
    Add Left of NumTree & Context: a type for a tree of numbers. Problem: add the numbers on the left branches. \\
    \hline
    Add Subway Station & Context: a subway system represented as a graph. Problem: add connections between stations. \\
    \hline
    Map Over Android Phones & Context: a type for phone models. Problem: a mapping function that only applies to Android phones. \\
    \hline
    Electrify Instruments & Context: a type hierarchy for several musical instruments with a flag that indicates if they are electric. Problem: mark instruments as electric. \\
    \hline
    Collatz Depth & Context: empty. Problem: requires a solution to see how far the given number is from converging on 1 in the Collatz sequence. \\
    \hline
    Mirror a Tree & Context: a type for trees. Problem: tree reflection. \\
    \hline
    Double Do It & Context: 2 of 3 helper functions. Problem:  define the third helper and the primary function. \\
    \hline
    Points and Lines & Context: types for point and lines types Problem: Manhattan distance. \\
    \hline
    Series & Context: a type that represents a series with its current value and update function. Problem: define a function that updates to the next value in the series. \\
    \hline
    Shapes & Context: several shape types and functions/methods. Problem: define a function that uses the helpers. \\
    \hline
    Social Time & Context: types that define calendar events. Problem: calculate time spent on a particular kind of event.  \\
    \hline
    Decode Message & Context: imports a common cryptography library. Problem: Decode AES encrypted message. \\
    \hline
    Verify Source & Context: imports a common cryptography library. Problem: verify signature. \\
    \hline
    Start AES & Context: imports a common cryptography library. Problem: generate an AES key and encrypts it. \\
    \hline
  \end{tabular}
\end{table}

\begin{figure}
\includegraphics[width=0.75\textwidth]{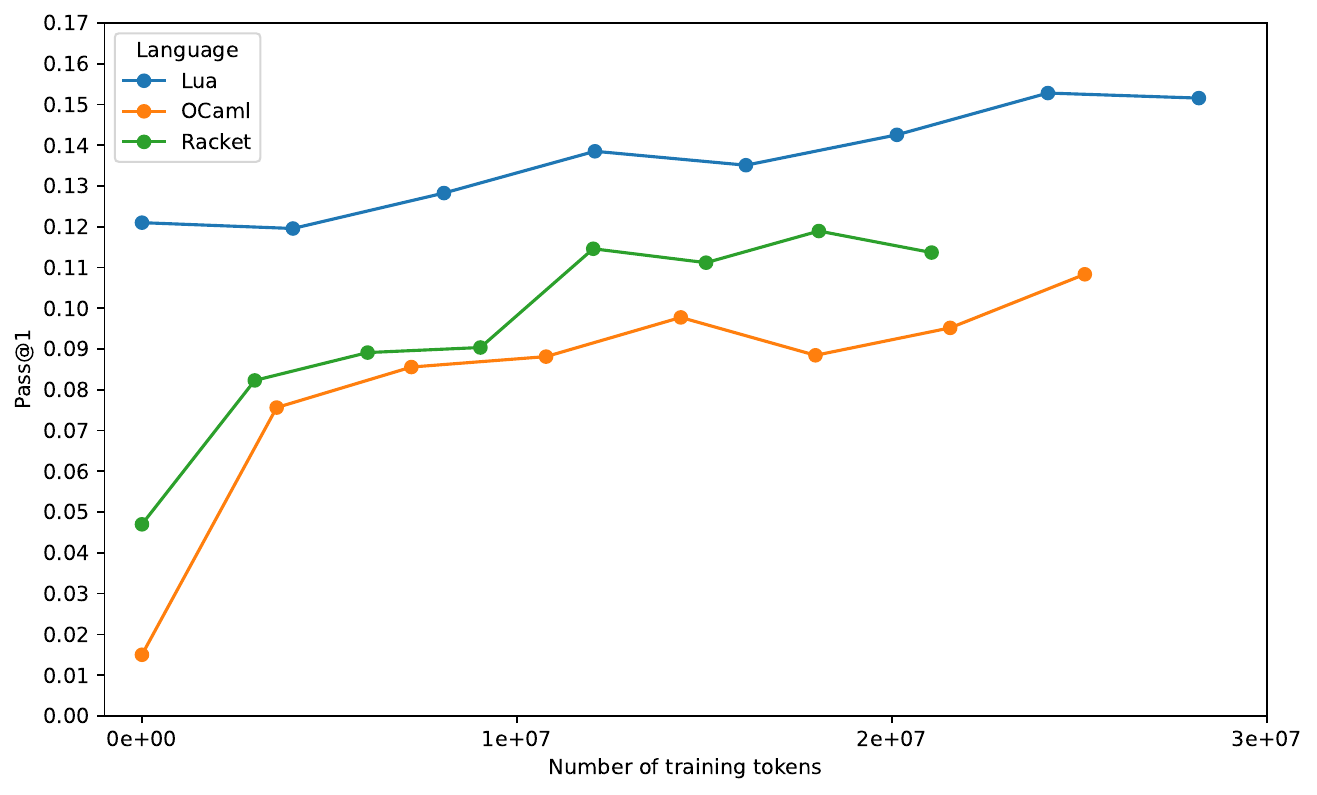}
\caption{We fine-tune three versions of StarCoderBase-1B on 25k \system{} generated training items. The $y$-axis measures performance on MultiPL-E and the $x$-axis counts the number of tokens. The points on the line mark epochs. Racket is far less verbose than Lua or OCaml, and thus has fewer tokens at each epoch. Epoch 0 is the base model.}
\label{fig:subset-1B}
\Description{}
\end{figure}

\subsubsection{Evaluating Coding Style}

\begin{table}
\caption{Our Racket style rubric for grading generated programs. The maximum score for a program is 15 points. Items are grouped into categories according to the type of error they entail.}
\label{fig:rkt-style-rubric}

    %cspell:disable
    \centering
    \begin{tabular}{|l|l|l|}
    \hline
    \textbf{Category} &  \textbf{Grading Item} & \textbf{Max Deduction} \\
    \hline
    \multirow{4}{*}{Text} & \multicolumn{1}{|l|}{Dangling parentheses} & \multicolumn{1}{|l|}{$-0.5$} \\\cline{2-3}
                          & \multicolumn{1}{|l|}{Line is too long} & \multicolumn{1}{|l|}{$-1$} \\\cline{2-3}
                          & \multicolumn{1}{|l|}{Using \texttt{car/cdr}} & \multicolumn{1}{|l|}{$-0.5$} \\\cline{2-3}
                         & \multicolumn{1}{|l|}{\texttt{cond} with round brackets} & \multicolumn{1}{|l|}{$-0.5$} \\\hline 
    \multirow{5}{*}{Definitions} & \multicolumn{1}{|l|}{\texttt{let} expression not at beginning of function body} & \multicolumn{1}{|l|}{$-1$} \\\cline{2-3}
                          & \multicolumn{1}{|l|}{Nesting \texttt{define} or \texttt{let} expressions} & \multicolumn{1}{|l|}{$-2$} \\\cline{2-3}
                          & \multicolumn{1}{|l|}{Unnecessary use of \texttt{let*} or \texttt{letrec}} & \multicolumn{1}{|l|}{$-1$} \\\cline{2-3}
                         & \multicolumn{1}{|l|}{Defining useless local variables} & \multicolumn{1}{|l|}{$-1$} \\\cline{2-3}
                        & \multicolumn{1}{|l|}{Not defining helpers or variables for reused code} & \multicolumn{1}{|l|}{$-1$} \\\hline
    \multirow{2}{*}{Conditionals} & \multicolumn{1}{|l|}{Nested \texttt{if} expression instead of \texttt{cond}} & \multicolumn{1}{|l|}{$-2$} \\\cline{2-3}
                          & \multicolumn{1}{|l|}{Using (\texttt{if COND \#t \#f})} & \multicolumn{1}{|l|}{$-1$} \\\hline
    Traversal & Using iteration when recursion is available & $-3$ \\
    \hline
    \end{tabular}
    %cspell:enable
\Description{}
\end{table}

\begin{figure}
\includegraphics[width=0.99\textwidth]{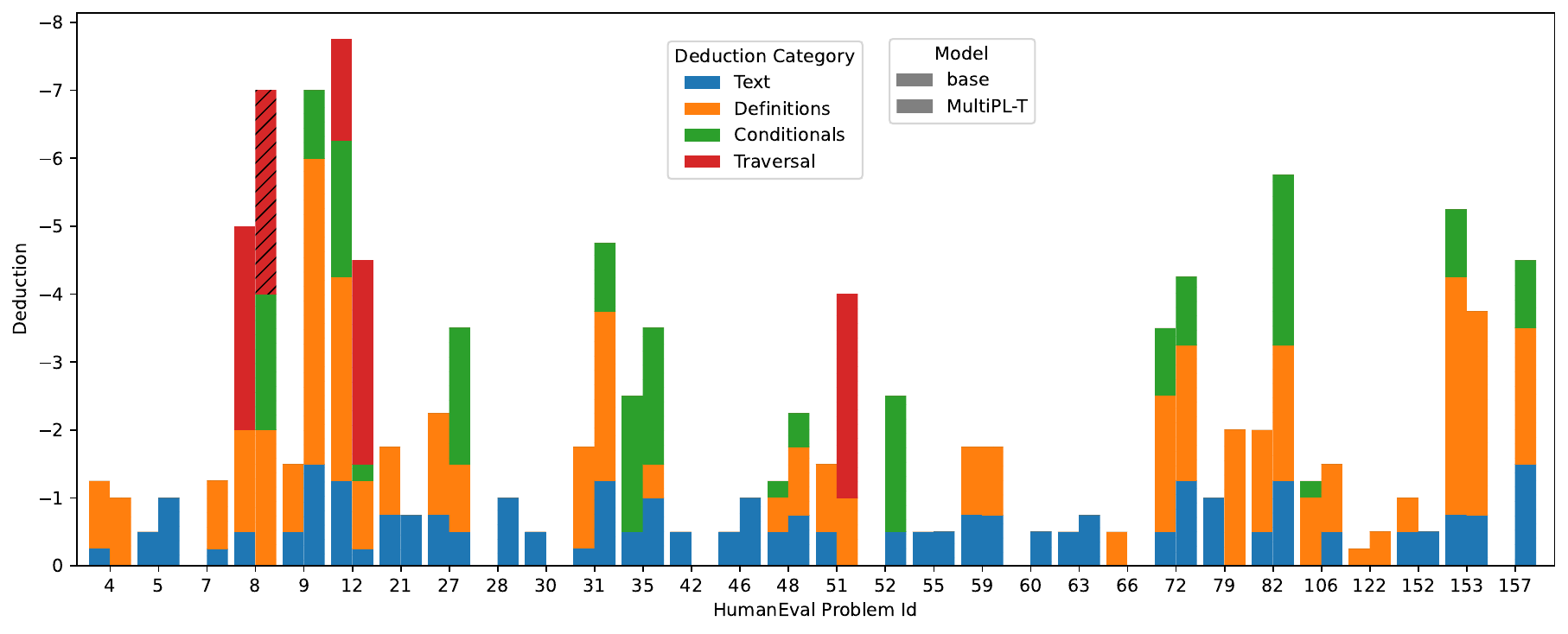}
\caption{Results of our Racket style evaluation for 30 out of 35 selected HumanEval problems. Omitted problems have no deductions and both models scored 15/15 points. Deductions for each item category are averaged across graders for each problem. Fine-tuning has a small negative impact on coding style.}
\label{fig:mean-qual-eval}
\Description{}
\end{figure}

\begin{figure}
\begin{subfigure}{\textwidth}
\begin{lstlisting}[style=Racket]
(define (rolling_max numbers)
  (define (max_in_list list)
	  (foldl max 0 list))
  (define (max_in_list_with_index list)
    #| body omitted |#)
	(let loop ((numbers numbers) (max_numbers '()))
		(if (empty? numbers)
			max_numbers
			(let* ((number (car numbers)) (max_number (max_in_list max_numbers)))
              (loop (cdr numbers) 
              (append max_numbers 
                      (list (if (> number max_number) number max_number))))))))    
\end{lstlisting}
\caption{A solution from the fine-tuned model. The \texttt{max\_in\_list\_with\_index} function is redundant.}
\end{subfigure}

\vspace{1em}
\begin{subfigure}{\textwidth}
\begin{lstlisting}[style=Racket]
(define (rolling_max numbers)
  (define (max_list numbers)
    (if (null? numbers)
        '()
        (cons (apply max numbers) (max_list (cdr numbers)))))
  (reverse (max_list (reverse numbers))))
\end{lstlisting}    
\caption{A better solution from the base model.}
\end{subfigure}
\caption{Two working solutions to the same HumanEval-Racket problem produced by the base and MultiPL-T fine-tuned model. In this case, the base model produces higher-quality code.}
\label{base-model-better}
\Description{}
\end{figure}

A potential limitation of \system{} is that it may negatively impact the style of generated code, since the training items are translated from Python. We study this issue in Racket using a qualitative evaluation process.
We developed a Racket style grading rubric (\Cref{fig:rkt-style-rubric}) based on the Racket style guide and our experience teaching and grading Racket programming assignments. The rubric outlines grading items and their corresponding deductions. Several deductions are designed for style problems that arise in code written by beginning students, such as needlessly long lines. Others penalize Lisp style, such as using \emph{car} and \emph{cdr} instead of \emph{first} and \emph{rest}. Finally, the largest deduction is for using imperative iteration when a simple functional solution is possible.

We use our rubric to grade the HumanEval solutions produced by StarCoderBase-1B before and after fine-tuning on \system{} Racket data. We grade only the 35 problems that both models can solve at least once (thus we omit several problems solvable only after fine-tuning). Since we generate several solutions for each problem, we select the most common working solution, and in the case of ties, we select the one with the best style. We grade 70 Racket programs in total: 35 produced by the base model and 35 after fine-tuning.
To avoid bias, we use two graders and anonymized the selection and grading process by assigning random IDs to each program and tracking their provenance on a hidden spreadsheet. The two graders have substantial Racket teaching experience and we find minimal discrepancy between them: their scores differed by more than 1 point for just 15 out of the 70 candidate programs, and only differed by 3--4 points for two programs.

We compute the mean overall score for the base model and \system{} model over the 35 HumanEval problems. We find that the base model achieves a style score of $89.5\%$ and the fine-tuned model $85.2\%$. In other words, the mean grade for a base model program is $13.4/15$ while for a fine-tuned model it is $12.8/15$. Thus fine-tuning leads to a slight decrease in our Racket style score.

% NOTE(arjun): Deliberately using \emph instead of \texttt for keywords below.
We inspect the 17 programs that scored higher for the base model than the fine-tuned model. We find that the fine-tuned model is more likely to use nested \emph{if} expressions in these programs, as well as performing iteration where recursion is available. Conversely, we inspect the eight  programs that scored higher in the \system{} model and find that the base model is more likely to use direct recursion with \emph{car/cdr} while the fine-tuned model uses Racket list abstractions. 
\cref{fig:mean-qual-eval} shows the breakdown of the kinds of deductions assigned per problem to each model, where deductions are averaged among graders.
\Cref{base-model-better} shows a graded pair where the base model scores better on coding style than the fine-tuned model, and 
\ifappendix \cref{fine-tuned-model-better} \else the supplementary material \fi shows a graded pair where the converse is true:
the fine-tuned model produces a cleaner solution than the base model.
Coding style aside, in both cases the lower quality solution is much more verbose than necessary.

The results of our evaluation are consistent with how we generated our training data from Python code. Overall, although \emph{fine-tuning slightly decreases the model's ability to generate idiomatic Racket code, it also significantly increases its ability to generate correct Racket code} as we showed earlier. Moreover, \emph{both models perform well on the Racket style rubric}, which suggests that the trade-off is minimal.

\begin{table}
\caption{The pass@50 performance of the basic prompt compared to the prompt with the 
original Python code for OCaml, Racket, and Lua.}
\label{tbl:prompt-ablation}  
\begin{tabular}{|l|r|r|}
\hline
Language & Basic & With Python Solution \\
\hline
OCaml & 26.1 & 23.9 \\
Racket & 34.7 & 56.8 \\
Lua & 51.4 & 68.5 \\
\hline
\end{tabular}
\end{table}
\subsection{\system{} Ablations}
\label{ablations}

The final part of our evaluation are ablations to demonstrate the efficacy of different parts of the \system{} pipeline. This section showcases results for a significant number of fine-tuned models. To manage costs, we use StarCoderBase-1B for the majority of the experiments.

\subsubsection{Fine-Tuning Efficiency}

\begin{wraptable}{r}{0.3\textwidth} % 'r' denotes right side, '0.3\textwidth' is the width of the table
\caption{Dataset sizes.}
\label{tab:datasetSizes}
\Description{}
\centering
\begin{tabular}{lr}
\toprule
Language & Size \\
\midrule
R      & 37,592 \\
Racket & 40,510 \\
OCaml  & 43,401 \\
Julia  & 45,000 \\
Lua    & 48,194 \\
\bottomrule
\end{tabular}
\end{wraptable}
We now investigate how the performance of StarCoderBase-1B on MultiPL-E varies during fine-tuning. On three languages (Lua, OCaml, and Racket), we fine-tune StarCoderBase-1B for seven epochs of \system{} data and evaluate at each epoch.
However, the datasets that we have generated are imbalanced in the number of training items (\Cref{tab:datasetSizes}).
To better balance the training sets, we randomly sample 25,000 items from each dataset to get similarly-sized fine-tuning sets for each language.\footnote{However, small differences remain: Lua is more verbose than OCaml, so 25,000 training items for Lua is more data than 25,000 training items for OCaml.}

In \Cref{fig:subset-1B}, we see that performance increases substantially after a single epoch of \system{} data for the lowest resource languages (Racket and OCaml). However, a higher-resource language (Lua) requires more data to realize even modest gains. These results are what one would expect: \emph{lower-resource languages enjoy easy and efficient gains from fine-tuning with \system{} data.}

On the other hand, \system{} requires more computing resources to generate data for low-resource languages, so the overall efficiency gap between languages is narrower than the figure suggests: it does not show the cost of generating training data. Nevertheless, \emph{\system{} data only needs to be generated once for a given language and can then be reused to train many models} (\cref{standard-eval}).

\subsubsection{Translation Versus Generation}
\label{sec:prompt-ablation}

\system{} uses StarCoderBase-15B to translate training items from Python to low-resource languages. The success rate of this translation is dependent on both the quality of the pretrained model (which is fixed) and the quality of the prompt (which we design).
We tried several prompt variations during development and eventually settled on a prompt that includes the original Python code in a comment (\Cref{sec:gen-target-funcs}).

Doing a complete ablation with all five languages and \textasciitilde133,000 functions would be prohibitively expensive. Instead, we run an experiment with a random sample of 1,000 source Python functions. We use the LLM to translate these 1,000 functions to OCaml, Racket, and Lua with two different prompt formats: with and without the original Python source. We generate 50 candidates for each prompt.\footnote{Thus this small experiment still requires 300,000 generations from the LLM and takes about 1/2 a day on an A100 GPU.}
We evaluate performance using pass@50, which is the likelihood that the model produces at least one correct solution in 50 attempts. 

As shown in \cref{tbl:prompt-ablation}, \emph{adding the original Python to the prompt substantially increases the likelihood of a successful translation to Racket and Lua, but slightly decreases the likelihood of a successful translation to OCaml}. We can only speculate about why this happens: Python may be misleading the model and OCaml seems further removed from Python than Racket or Lua.

\begin{table}
\caption{Without test-based validation, the LLM-translated training items are poor quality and fine-tuning is not effective. The performance of the 15B model decreases slightly compared to the base model. The performance of the 1B model increases slightly, but is far worse than the result we get with the tested items.}
\label{table:no-tests}  
\centering
\begin{tabular}{|l|r|r|r|}
\hline
Model & Baseline & \system{} & No Validation \\
\hline
StarCoderBase-1B & 4.7 & 11.3 & 8.6 \\
StarCoderBase-15B & 11.8 & 21.0 & 11.6 \\
\hline
\end{tabular}
\end{table}

\subsubsection{Testing Is Critical for \system{}}

\system{} relies on tests to validate the LLM-translated training items. We reason that since we are translating to programming languages on which the LLM performs poorly at synthesis, it is also likely to perform poorly at translation. But, to validate this claim, we fine-tune models on Racket data without
test validation. We get poor results as expected (\cref{table:no-tests}). This shows that the effort we take to generate, validate, and compile tests is essential for \system{}.

\subsubsection{The Impact of Deduplication}

Prior work has shown that data duplication decreases performance while increasing training time~\cite{lee2022deduplicating,allal:santacoder}. To demonstrate its impact in our work, we do an experiment on the dataset where it has the most impact---where deduplication discards the most data---which is for Lua. This is to be expected because the number of nearly duplicate functions will be higher when the data generating LLM translates functions with a higher success rate. Of the languages we target, our data generating LLM (StarCoderBase15-B) has the highest pass rate on Lua.

Before deduplication, the Lua dataset has 1.4M functions, but after deduplication we are left with 48K (reported in \cref{tab:datasetSizes}). For this experiment, we fine-tune StarCoderBase-1B on a single H100 GPU on the undeduplicated dataset, with exactly the same hyperparameters we used earlier (\cref{evaluation}). We find the following:

\begin{enumerate}
    
    \item Without deduplication, the fine-tuned StarCoderBase-1B gets a \textbf{17.1} pass@1 score on MultiPL-E and takes \textbf{12 hours} to train.

    \item With deduplication, the fine-tuned StarCoderBase-1B gets \textbf{17.3} pass@1 on MultiPL-E and takes less than \textbf{30 minutes} to train.

\end{enumerate}
The two pass@1 scores are very close. But, without deduplication, training takes significantly longer, because of the increase in dataset size.

The impact would be more significant with a larger model. For example, it took us 8 hours on 8xH100 GPUs to fine-tune StarCoderBase-15B with a deduplicated dataset. Without deduplication, fine-tuning would take several days.  Similarly, fine-tuning the 33B and 70B models would take even longer. Thus deduplication is a way to manage the time and cost of training large models.

\subsubsection{Targeting the Source Language}

The complete \system{} toolchain is designed to produce training data for a target low-resource language from a source language (Python). However, it is also possible to directly evaluate the first part of the \system{} toolchain, which builds a high-quality dataset for the source language (Python). In this section, we fine-tune select LLMs on Python using the dataset of 133,668 Python functions that are documented, tested, typeable, and filtered on test coverage (\cref{sec:py-source-funcs,generating-unit-tests}).

We fine-tune and evaluate StarCoderBase-1B directly on this dataset of Python functions. The base model gets a \textbf{15.1} pass@1 score on Python, whereas the fine-tuned model gets \textbf{15.0} pass@1.\footnote{We observe this score on epoch 3. On epochs 1--7, the score varies between 10.7 and 15.0.} Thus, it is clear that fine-tuning directly on this subset of Python does not produce the same improvements that we see on the low-resource languages after translation.

We hypothesize that this occurs for the following reason: StarCoderBase-1B has already been trained on three epochs of this Python data. We are not fine-tuning the model on novel data, thus there is little new for the model to learn, and this is reflected in the unchanged pass@1 score. In contrast, when we translate this data to a low-resource language using the full \system{} pipeline, we are effectively creating novel data for the model to learn.

\section{Discussion}

We have shown that \system{} is an effective and efficient method for generating semi-synthetic training data for low-resource programming languages. In this section, we discuss the implications of extending \system{} in various ways.

\paragraph{Generalizing to other programming languages}

We hope it is clear to the reader that the \system{} approach is straightforward to generalize to more programming languages.
The language-specific work involves 1)~translating comments and function signatures into an appropriate prompt, and 2)~writing a compiler that can translate simple assertions from the source to the target. MultiPL-E already supports both these steps for 20+ programming languages, several of which are low-resource, including D, Bash, MATLAB, Haskell, and Perl. So, generating fine-tuning sets for these languages may just be a matter of running the \system{} pipeline for a few days on GPUs.

Some combinations of source and target languages may be more effective than others. For example, consider building semi-synthetic training data for Rust. We could directly leverage the datasets in this paper and utilize a Code LLM to translate Python to Rust. However, we speculate that it would be more effective to modify \system{} to support a source language closer to Rust, such as C++.

\paragraph{Generalizing to other LLMs}
Although this paper focuses on fine-tunes of the Code Llama and StarCoder family of Code LLMs, our datasets could also be used to fine-tune other LLMs. We fine-tune DeepSeek Coder and StarCoder2 on Racket, and we expect our results will generalize to the other languages.

\paragraph{Regenerating \system{} data}
The \system{} pipeline in this paper uses StarCoderBase-15B. It should be clear that using a more capable model would improve conversion rates, and thus produce better results. A proprietary model such as GPT-4 would likely produce the best results, but doing so would violate its terms of use~\cite{openai2023terms}. An unusual result in this paper is that \system{} can use a weaker model (StarCoderBase-15B) to improve much better models (CodeLlama-70B, DeepSeekCoder-34B, and StarCoder2-15B) on low-resource languages.

\paragraph{``No-resource'' languages}

\system{} targets low-resource languages, but it is unlikely to work as-is for ``no-resource'' languages that StarCoderBase is not trained on at all.
It may be possible to cleverly prompt the model with enough information about a no-resource language to bootstrap data generation. But, doing so efficiently may be challenging.

\paragraph{Composability with self-instruct}

Although we have argued that self-instruct is unlikely to succeed on a low-resource language, self-instruct and \system{} could be composed together in a natural way: one could generate a high-quality dataset of instructions in a high-resource language, and then use \system{} to translate them to a low-resource language. Given the effectiveness of WizardCoder~\cite{ziyang:wizard-coder} and Magicoder~\cite{wei:magicoder} at Python, this composition seems likely to succeed.

\paragraph{Other kinds of benchmarks}

A lot of effort has been put into evaluating the Python programming abilities of LLMs, but there are far fewer benchmarks for low-resource languages. For example, there are Python benchmarks for specialized tasks, such as data science~\citep{lai:ds1000}, and with prompts authored by specific populations, such as beginning programmers~\citep{babe:studenteval}. We need to develop these kinds of benchmarks for low-resource languages to truly understand the capabilities and limitations of Code LLMs.

% NOTE(arjun): Related Work is immediately before Conclusion, which is typical PL style.
\section{Related Work}

\paragraph{Code translation with language models and unit tests}

A number of projects use language models to translate code between programming languages and test that the generated translations are correct by compiling working unit tests from one language to another.
TransCoder-ST~\cite{roziere:transcoder-st} and CMTrans~\cite{cmtrans} use these techniques to generate training data for a code translation model between Java, Python, and C++, whereas MultiPL-E~\cite{cassano:multipl-e}, MBXP~\cite{athiwaratkun:mbxp}, and BabelCode~\cite{babelcode} translate Code LLM benchmarks from Python to 10+ programming languages.
A distinguishing feature of \system{} is that it employs an off-the-shelf pretrained Code LLM (StarCoder) to both generate test cases and translate code to low-resource languages.
When the aforementioned papers were written, the best open Code LLMs were far less capable than StarCoder: they were trained on fewer programming languages using far less training data, and they were an order of magnitude smaller. Thus we believe the \system{} approach would have likely failed.
Although capable closed models were available, they were either rate-limited or prohibitively expensive for the scale of data generation that \system{} needs. For example, \citet{cassano:multipl-e} 
report that they used a commercial model during a free beta period, but it would have cost \$37,000 with equivalent released models.
\system{}'s data generation requires an order of magnitude more queries, making it prohibitively expensive to use with commercial models at 2023 prices.

\paragraph{Multi-lingual language models of code}
Training models on a dataset of code written in programming languages similar to the target language, known as transfer learning, is a commonly studied method for improving model performance on a specific programming language \cite{multilingual-training,baltaji2024learning,language-selection}.
\citet{chen2022transferability} explore this method in the context of low-resource languages by pretraining and fine-tuning small encoder-only Transformers with different programming languages.
They then evaluate their performance on code search and summarization tasks for Ruby, a language with a dataset 10x smaller than that of Python.
Their findings suggest that multilingual pretraining can improve performance at generating code in low-resource languages.
However, this strategy may falter with languages that significantly differ in syntax or semantics, such as Racket.
For example, \citet{baltaji2024learning} employ a similar methodology to train models on 41 different programming language pairs, noting that Scheme, the predecessor of Racket, benefits less from transfer learning compared to other languages.
In such scenarios, data generation strategies like \system{} are deemed more likely to achieve success.
Moreover, these studies predominantly rely on the BLEU score and other syntax-based metrics for evaluating correctness, which have been criticized for their inadequacy in code generation tasks~\cite{chen2021evaluating}.
In our work, we utilize test-based validation of 
both our training data and evaluation metrics to ensure the correctness of generated code,
which is a more reliable measure of model performance.

Our work begins with the observation that Code LLM performance can vary significantly by language, and this has been observed repeated in prior work. Most closely related to \system{} are benchmarks for the natural language to code task~\cite{cassano:multipl-e,babelcode,athiwaratkun:mbxp}. However, similar trends occur when Code LLMs are used for other tasks, such as fuzzing and translation~\cite{xia:universal-fuzzing,pan:lost-in-translation}.

\paragraph{Instruction tuning}

To get an LLM to perform a desired task, the user must prompt it in the right way. There are several techniques for \emph{instruction tuning} LLMs to better follow natural, human-written instructions.
One approach uses human annotators to write sample instructions and give feedback on a large number of model generations~\cite{ouyang:instructgpt}, but this is expensive and requires significant resources.
A cheaper approach is to have a capable LLM \emph{self-instruct} to generate instructions from a relatively small set of human-written seed instructions~\cite{wang:self-instruct}.
Evol-Instruct uses an LLM to create variations of instructions~\cite{ziyang:wizard-coder}.
These techniques have been used to create datasets for instruction-tuning Code LLMs~\cite{codealpaca,muennighoff:octopack,ziyang:wizard-coder,wei:magicoder}.
These datasets focus on high-resource languages, and, as we show in \cref{use-self-instruct}, they are unlikely to succeed for low-resource languages.

\paragraph{Training on high-quality data}

Training on high-quality data is an effective way to reduce both the size of an LLM and the volume of training data needed, while maintaining performance.
\citet{gunasekar2023textbooks} achieves high HumanEval scores on a small model with a modest amount of ``textbook quality'' training data. This includes both natural and synthetic data generated by a more capable model.
Their work targets Python, and we argue in \cref{alternatives} that the approach is less likely to succeed with low-resource languages.

\paragraph{Proprietary Code LLMs}

At the time of writing, there are proprietary LLMs that perform better at programming tasks than the open models we build upon \citep{openai2023gpt4,anthropic2023claude,anil2023palm}. However, most of these models only support inference (i.e., running the trained model) and not training or fine-tuning.
Even when fine-tuning is possible, because these models are trained on closed data sets, we would not be able to compare \system{} to the natural baseline of training longer on existing data.
Moreover, a significant limitation arises from the proprietary licensing constraints of these models. Many of their licenses expressly forbid the use of generated data to train other models \cite{openai2023terms,google2023terms,anthropic2023terms}.

%\paragraph{Models fine-tuned on the output of proprietary Code LLMs}

Even though, many proprietary models forbid using their generated data to train other models, there are models that do so. These models are \emph{distillations} of much larger proprietary models, and it is not possible to compare them to the natural baseline, which is training longer on existing data, which for these models is the proprietary OpenAI training set. The strength of the \system{} approach is not that it does better in an absolute sense, but that fine-tuning with \system{} data is better than training longer on existing data. We show that this holds for StarCoderBase  and its training set, and can only do so because both are open.

\section{Conclusion}

In the last few years, Code LLMs have rapidly made their way into more and more programming tools.
However, the quality and reliability of Code LLMs is highly language-dependent: they can be remarkable on high-resource programming languages, but are far less impressive at working with low-resource languages~\citep{cassano:multipl-e}.
It is possible that in the near future, a large number of programmers will expect LLM-based technology to just work, just as many programmers today expect syntax highlighting, continuous analysis, or type-based completion in their programming environments.
We hope that \system{}---a methodology for generating large-scale, high-quality fine-tuning datasets in low-resource languages---will help low-resource languages compete in a world where many developer tools rely on Code LLMs.

The \system{} fine-tuning data (and code) are also open: they are constructed from the StarCoder training data (The Stack) and augmented by StarCoder itself. We deliberately do not use a more capable proprietary model to fine-tune an open model. This allows us to demonstrate that fine-tuning on \system{} data is more effective and efficient than training longer on existing data. We evaluate \system{} in several other ways, including a new  benchmark that exercises in-context learning and a qualitative evaluation of coding style. When compared to other fine-tunes of the same base model, \system{} achieves state-of-the-art results for \alllangs{}.

% NOTE(arjun): Instructions: To help readers find data and software, OOPSLA recommends adding a section just before the references titled Data-Availability Statement. If the paper has an artifact, cite it here. If there is no artifact, this section can explain how to obtain relevant code. The statement does not count toward the OOPSLA 2023 page limit. It may be included in the submitted paper; in fact we encourage this, even if the DOI is not ready yet.

\section*{Data-Availability Statement}

All code, data, and models from this paper are available with open licenses. \emph{Code:} available on GitHub and archived on Zenodo; \emph{Datasets:} available on Hugging Face; and \emph{Models:} available on Hugging Face. In particular:

\begin{enumerate}
    \item A guide to reproduce the results in this article is available at \href{https://doi.org/10.5281/zenodo.12453932}{doi.org/10.5281/zenodo.12453932}.

    \item The \system{} datasets are available at \href{https://doi.org/10.57967/hf/2941}{doi.org/10.57967/hf/2941} with links to several models fine-tuned on these datasets.

\end{enumerate}

\section*{Acknowledgements}

We thank Loubna Ben Allal, Harm de Vries, Joydeep Biswas, Matthias Felleisen, Shriram Krishnamurthi, and Leandro von Werra for helpful conversations. Thanks to Leandro von Werra for including the \system{} datasets in the StarCoder2 training corpus. We also thank the OOPSLA reviewers for their feedback. We especially thank the artifact evaluation committee for their patience reviewing our complex artifact, as well as the AEC chairs (Guillaume Baudart and Sankha Narayan Guria) for facilitating review. This work used computing resources provided by Northeastern Research Computing, New England Research Cloud, and Joydeep Biswas (UT Austin). Federico Cassano was affiliated with Roblox for most of his work on this paper. This work is partially supported by the National Science Foundation (SES-2326173, SES-2326174, and SES-2326175).

\bibliographystyle{ACM-Reference-Format}
\bibliography{main, arxiv_and_websites}

\end{document}